\documentclass[prd,twocolumn,aps,amsmath,amssymb,nofootinbib,preprintnumbers,fleqn]
{revtex4}

\usepackage{graphicx}
\usepackage{dcolumn}
\usepackage{bm}
\usepackage{amsmath}
\usepackage{amsfonts}
\usepackage{subfigure}

\def\ls{\mathrel{\lower4pt\vbox{\lineskip=0pt\baselineskip=0pt
           \hbox{$<$}\hbox{$\sim$}}}}
\def\gs{\mathrel{\lower4pt\vbox{\lineskip=0pt\baselineskip=0pt
           \hbox{$>$}\hbox{$\sim$}}}}
\def\drawbox#1#2{\hrule height#2pt

\hbox{\vrule width#2pt height#1pt \kern#1pt
              \vrule width#2pt}
              \hrule height#2pt}

\def\Asym#1#2{\vcenter{\vbox{\drawbox{#1}{#2}
              \kern-#2pt       
              \drawbox{#1}{#2}}}}

\def\half{\frac{1}{2}}

\newcommand{\beq}{\begin{equation}}
\newcommand{\eeq}{\end{equation}}

\begin{document}

\title{A Gravitational Wave Background from Reheating after Hybrid Inflation}

\author{Juan Garc\'\i a-Bellido, Daniel G. Figueroa and Alfonso Sastre}
\affiliation{Departamento de F\'\i sica Te\'orica, 
Universidad Aut\'onoma de Madrid, Cantoblanco, 28049 Madrid, Spain\\
Instituto de F\'\i sica Te\'orica CSIC-UAM, 
Universidad Aut\'onoma de Madrid, Cantoblanco, 28049 Madrid, Spain}

\date{November 26th, 2007}

\begin{abstract}
The reheating of the universe after hybrid inflation proceeds through
the nucleation and subsequent collision of large concentrations of
energy density in the form of bubble-like structures moving at
relativistic speeds. This generates a significant fraction of energy
in the form of a stochastic background of gravitational waves, whose
time evolution is determined by the successive stages of
reheating: First, tachyonic preheating makes the amplitude of gravity
waves grow exponentially fast. Second, bubble collisions add a new
burst of gravitational radiation. 
Third, turbulent motions finally sets the end of gravitational waves 
production. From then on, these waves propagate unimpeded to us.
We find that the fraction of energy density today in these
primordial gravitational waves could be significant for GUT-scale
models of inflation, although well beyond the frequency range
sensitivity of gravitational wave observatories like LIGO, LISA or
BBO. However, low-scale models could still produce a detectable signal
at frequencies accessible to BBO or DECIGO. For comparison, we have
also computed the analogous gravitational wave background from some 
chaotic inflation models and obtained
results similar to those found by other groups. The discovery of such
a background would open a new observational window into the very early
universe, where the details of the process of reheating, i.e. the Big
Bang, could be explored.  Moreover, it could also serve in the future
as a new experimental tool for testing the Inflationary Paradigm.
\end{abstract}
\preprint{IFT-UAM/CSIC-07-38}
\maketitle


\section{Introduction}

Gravitational waves (GW) are ripples in space-time that travel at the
speed of light, and whose emission by relativistic bodies represents a
robust prediction of General Relativity.  The change in the orbital
period of a binary pulsar known as PSR 1913+16 was used by Hulse and
Taylor~\cite{HulseTaylor} to obtain indirect evidence of their
existence. Although gravitational radiation has not been directly
detected yet, it is expected that the present universe should be
permeated by a diffuse background of GW of either an astrophysical or
cosmological origin~\cite{Maggiore}. Astrophysical sources, like the
gravitational collapse of supernovae or the neutron star and black
hole binaries' coalescence, produce a stochastic gravitational wave
background (GWB) which can be understood as coming from unresolved
point sources. On the other hand, among the backgrounds of
cosmological origin, we find the approximately scale-invariant
background produced during inflation~\cite{Starobinsky}, or the GWB
generated at hypothetical early universe thermal phase transitions,
from relativistic motions of turbulent plasmas or from the decay of
cosmic strings~\cite{Maggiore}. Fortunately, these backgrounds have
very different spectral shapes and amplitudes that might, in the
future, allow gravitational wave observatories like the Laser
Interferometer Gravitational Wave Observatory (LIGO)~\cite{LIGO}, the
Laser Interferometer Space Antenna (LISA)~\cite{LISA}, the Big Bang
Observer (BBO)~\cite{BBO} or the Decihertz Interferometer
Gravitational Wave Observatory (DECIGO)~\cite{DECIGO}, to disentangle
their origin~\cite{Maggiore}. Unfortunately, due to the weakness of
gravity, this task will be extremely difficult, requiring a very high
accuracy in order to distinguish one background from another. It is
thus important to characterize as many different sources of GW as
possible.

There are, indeed, a series of constraints on some of these
backgrounds, the most stringent one coming from the large-scale
polarization anisotropies in the Cosmic Microwave Background (CMB),
which may soon be measured by Planck~\cite{Planck}, if the scale of
inflation is sufficiently high. There are also constraints coming from
Big Bang nucleosynthesis~\cite{BBN}, since such a background would
contribute as a relativistic species to the expansion of the universe
and thus increase the light element abundance. There is also a
constraint coming from millisecond pulsar timing~\cite{pulsar}.
Furthermore, it has recently been proposed a new constraint on a GWB
coming from CMB anisotropies~\cite{Elena}. Most of these constraints
come at very low frequencies, typically from $10^{-18}$ Hz to
$10^{-8}$ Hz, while present GW detectors work at frequencies of order
1-100 Hz, and planned observatories will range from $10^{-3}$ Hz of
LISA to $10^{3}$ Hz of Advanced-LIGO~\cite{Maggiore,LIGO}. If early
universe phenomena like first order phase
transitions~\cite{KosowskyTurner,Nicolis} or cosmic
turbulence~\cite{Turbulence} occurred around the electro-weak (EW)
scale, there is a chance than the GW detectors will measure the
corresponding associated backgrounds. However, if such early universe
processes occurred at the GUT scale, their corresponding backgrounds
will go undetected by the actual detectors, since these cannot reach
the required sensitivity in the high frequency range of $10^7-10^9$
Hz, corresponding to the size of the causal horizon at that time.
There are however recent proposals to cover this
range~\cite{Nishizawa2007}, which may become competitive in the near
future.

Moreover, present observations of the CMB anisotropies and
the Large Scale Structure (LSS) distribution of matter seem to suggest
that something like Inflation must have occurred in the very early
universe. We ignore what drove inflation and at what scale it took
place. However, approximately scale-invariant density perturbations,
sourced by quantum fluctuations during inflation, seem to be the most
satisfying explanation for the CMB anisotropies. Together with such
scalar perturbations one also expects tensor perturbations (GW) to be
produced, with an almost scale-free power
spectra~\cite{Starobinsky}. Because of the weakness of gravity, this
primordial inflationary GWB should decouple from the rest of matter as
soon as it is produced, and move freely through the Universe till
today. At present, the biggest efforts employed in the search for
these primordial GW come from the indirect effect that this
background has on the B-mode polarization anisotropies of the
CMB~\cite{Planck}, rather than via direct detection. The detection of
such a background is crucial for early universe cosmology because it
would help to determine the absolute energy scale of inflation, a
quantity that for the moment is still uncertain, and would open the
exploration of physics at very high energies.

In the early universe, after inflation, other backgrounds of GW could
have been produced at shorter wavelengths, in a more 'classical'
manner, rather than sourced by quantum fluctuations. In particular,
whenever there are large and fast moving inhomogeneities in the matter
distribution, one expects the emission of GW.  This is much like the
situation in classical electrodynamics, but with some differences. At
large distances from the source, the amplitude of the electromagnetic
field $A_i$ is expressed as the first derivative of the dipole moment
$d_i$ of the charge distribution of the source, $A_i\simeq \dot d_i /
cr$, while the amplitude of the GW is given by the second derivative
of the quadrupole moment of the mass distribution, $h_{ij} \simeq
G\ddot Q_{ij} / c^4r$. In both cases, the larger the velocity of the
matter/charge distribution, the larger the amplitude of the radiation
produced. Nevertheless, the main difference between the two cases is
the weakness of the strength of gravity to that of
electromagnetism. Thus, in order to produce a significant amount of
gravitational radiation, it is required that the motion of huge masses
occurs at speeds close to that of light for the case of astrophysical
sources, or a very fast motion and high density contrasts in the
continuous matter distribution for the case of cosmological sources.
In fact, this is believed to be the situation at the end of inflaton,
during the conversion of the huge energy density driving inflation
into radiation and matter at the so-called {\it reheating} of the
Universe~\cite{preheating}. Such an event corresponds to 
the actual Big Bang of the Standard Cosmological Model.

Note that any background of GW coming from the early universe, if
generated below Planck scale, immediately decoupled upon production,
as can be easily understood by the following dimensional analysis
argument. Assuming that gravitons were in thermal equilibrium with the
early universe plasma, at a temperature $T$, the gravitons' cross
section should be of order $\sigma\sim G^2T^2$. Then, given the
graviton number density $n\sim T^3$ and velocity $v = 1$, the
gravitons' interaction rate should be $\Gamma = \left\langle n\sigma
v\right\rangle\,\sim\,T^5/M_p^4$. Since the Hubble rate is $H\sim
T^2/M_p$, then $\Gamma\,\sim\,H\,(T/M_p)^3$, so gravitons could {\it
not} be kept in equilibrium with the surrounding plasma for
$T\,<\,M_p$. Therefore, GW produced well after Planck scale will
always be decoupled from the plasma, and whatever their spectral
signatures, they will retain their shape throughout the expansion of
the Universe. Thus, the characteristic frequency and shape of the GWB
generated at a given time should contain information about the very
early state of the Universe in which they were produced.  Actually, it
is conceivable that, in the not so far future, the detection of these
GW backgrounds could be the only way we may have to infer the physical
conditions of the Universe at such high energy scales, which certainly
no particle collider will ever reach. However, the same reason that
makes GW ideal probes of the early universe $-$ the weakness of gravity
$-$, is responsible for the extreme difficulties we have for their
detection on Earth. For an extensive discussion see Ref.~\cite{Thorne}.

In a recent letter~\cite{GBF} we described the stochastic background
predicted to arise from reheating after hybrid inflation. In this
paper we study in detail the various processes involved in
the production of such a background, whose detection could open a new
window into the very early universe. In the future, this background
could also serve as a new tool to discriminate among different
inflationary models, as each of these would give rise to a different
GWB with very characteristic spectral features. The first stage of the
energy conversion at the end of inflation,
preheating~\cite{preheating}, is known to be explosive and extremely
violent, and quite often generates in less than a Hubble time the huge
entropy measured today. The details of the dynamics of preheating
depend very much on the model and are often very complicated because
of the non-linear, non-perturbative and out-of-equilibrium character
of the process itself. However, all the cases have in common that only
specific resonance bands of the fields suffer an exponential
instability, which makes their occupation numbers grow by many orders
of magnitude. The shape and size of the spectral bands depend very
much on the inflationary model. If one translates this picture into
position-space, the highly populated modes correspond to large
time-dependent inhomogeneities in the matter distributions which
acts, in fact, as the source of GW we are looking for.

For example, in single field chaotic inflation models, the coherent
oscillations of the inflaton during preheating generates, via
parametric resonance, a population of highly occupied modes that
behave like waves of matter. They collide among themselves and their
scattering leads to homogenization and local thermal equilibrium.
These collisions occur in a highly relativistic and very asymmetric
way, being responsible for the generation of a stochastic
GWB~\cite{TkachevGW,JuanGW,EastherLim} with a typical frequency today
of the order of $10^{7} - 10^{9}$ Hz, corresponding to the present
size of the causal horizon at the end of high-scale inflation. There
is at present a couple of experiments searching for such a background,
see Refs.~\cite{Nishizawa2007}, based of laser interferometry, as well
as by resonant superconducting microwave cavities~\cite{Picasso}.

However, there are models like hybrid inflation in which the end of
inflation is sudden~\cite{hybrid} and the conversion into radiation
occurs almost instantaneously. Indeed, since the work of
Ref.~\cite{tachyonic} we know that hybrid models preheat in an even
more violent way than chaotic inflation models, via the spinodal
instability of the symmetry breaking field that triggers the end of
inflation, irrespective of the couplings that this field may have to
the rest of matter. Such a process is known as \textit{tachyonic
preheating}~\cite{tachyonic,symmbreak} and could be responsible for
copious production of dark matter particles~\cite{ester}, lepto and
baryogenesis~\cite{CEWB}, topological defects~\cite{tachyonic},
primordial magnetic fields~\cite{magnetic}, etc.

It was speculated in Ref.~\cite{JuanGW} that in (low-scale) models of
hybrid inflation it might be possible to generate a stochastic GWB in
the frequency range accessible to present detectors, if the scale of
inflation was as low as $H_{\rm inf} \sim 1$ TeV. However, the
amplitude was estimated using the parametric resonance formalism of
chaotic preheating, which may not be applicable in this case. In
Ref.~\cite{symmbreak} (from now on referred to as paper I), it was
shown that the process of symmetry breaking proceeds via the
nucleation of dense bubble-like structures moving at the speed of
light, which collide and break up into smaller structures (see Figs.~7
and~8 of paper I). We conjectured at that time that such collisions
would be a very strong source of GW, analogous to the GW production
associated with strongly first order phase
transitions~\cite{KosowskyTurner}. As we will show in this paper, this
is indeed the case during the nucleation, collision and subsequent
rescattering of the initial bubble-like structures produced after
hybrid inflation. During the different stages of reheating in this
model, gravity waves are generated and amplified until the Universe
finally thermalizes and enters into the initial radiation era of the
Standard Model of Cosmology.  From that moment until now, during the
whole thermal history of the expansion of the universe, this cosmic
GWB will be redshifted as a radiation-like fluid, totally decoupled
from any other energy-matter content of the universe, such that
today's ratio of energy stored in these GW to that in radiation, could
range from $\Omega_{_{\rm GW}}h^2\sim 10^{-8}$, peacked around $f\sim
10^7$~Hz for the high-scale models, to $\Omega_{_{\rm GW}}h^2\sim
10^{-11}$, peacked around $f\sim 1$ Hz for the low-scale models.

Finally, let us mention that since the first paper by Khlebnikov and
Tkachev ~\cite{TkachevGW}, studing the GWB produced at reheating after
chaotic inflation, it seems appropriate to reanalyze this topic in a
more detailed way. The idea was extended to hybrid inflation
in~\cite{JuanGW}, but within the parametric resonance formalism. It was
also revisited very recently in Ref.~\cite{EastherLim,EastherLim2} for
the $\lambda\phi^4$ and $m^2\phi^2$ chaotic scenarios, and reanalysed
again for hybrid inflation in Ref.~\cite{GBF}, this time using the new
formalism of tachyonic preheating~\cite{tachyonic,symmbreak}.  Because
of the increase in computer power of the last few years, we are now
able to perform precise simulations of the reheating process in a
reasonable time scale. Moreover, understanding of reheating has
improved, while gravitational waves detectors are beginning to attain
the aimed sensitivity~\cite{LIGO}. Furthermore, since these cosmic
GWBs could serve as a deep probe into the very early universe, we
should characterize in the most detailed way the information that we
will be able to extract from them.

The paper is divided as follows. In Section II we briefly review the
hybrid model of inflation. Section III is dedicated to our approach
for extracting the power spectrum of GW from reheating. In section
IV, we give a detailed account of the lattice simulations performed
with two codes: our own FORTRAN parallelized computer code (running in
the MareNostrum supercomputer~\cite{BSC} and in our UAM-IFT
cluster~\cite{IFTcluster}), as well as with a modified version of the
publicly available C++ package LATTICEEASY \cite{LatticeEasy}. Section
V is dedicated to study the spatial distribution of the
production of gravitational waves. In Section VI, we reproduce as a
crosscheck, some of the results of~\cite{TkachevGW,EastherLim,DufauxGW} 
concerning the GWB produced at reheating after chaotic inflation models. 
Finally, in section VII, we give our conclusions and perspectives for 
the future.

\section{The Hybrid Model}

Hybrid inflation models~\cite{hybrid} arise in theories of particle
physics with symmetry breaking fields ('Higgses') coupled to flat
directions, and are present in many extensions of the Standard Model,
both in string theory and in supersymmetric theories~\cite{LythRep}. 
The potential in these models is given by
\begin{equation}\label{higgsPotential}
\indent V(\Phi,\chi) = \lambda\left(\Phi^{\dagger}\Phi - 
{v^2\over2}\right)^2 + g^2\chi^2\Phi^{\dagger}\Phi 
+ \frac{1}{2}\mu^2\chi^2\,,
\end{equation}
where the contraction $\Phi^{\dagger}\Phi$ should be understood as the
trace ${\rm Tr}\,\Phi^{\dagger}\Phi=\half|\phi|^2$.  Inflation occurs
along the lifted flat direction, satisfying the slow-roll conditions
thanks to a large vacuum energy $\rho_0=\lambda v^4/4$. Inflation ends
when the inflaton $\chi$ falls below a critical value and the symmetry
breaking field $\phi$ acquires a negative mass squared, which triggers
the breaking of the symmetry and ends in the true vacuum, $|\phi|=v$,
within a Hubble time. These models do not require small couplings in
order to generate the observed CMB anisotropies; e.g. a working model
with GUT scale symmetry breaking, $v=10^{-3}\,M_P$, with a Higgs
self-coupling $\lambda$ and a Higgs-inflaton coupling $g$ given by
$g=\sqrt{2\lambda}=0.05$, satisfies all CMB constraints~\cite{WMAP},
and predicts a tiny tensor contribution to the CMB polarization. The
main advantage of hybrid models is that, while most chaotic inflation
models can only occur at high scales, with Planck scale values for the
inflaton, and $V_{\rm inf}^{1/4} \sim 10^{16}$ GeV, one can choose the
scale of inflation in hybrid models to range from GUT scales all the
way down to GeV scales, while the Higgs \textit{v.e.v.} can range from
Planck scale, $v=M_P$, to the Electroweak scale, $v=246$ GeV, see
Ref.~\cite{hybrid,CEWB}.

There are a series of constraints that a hybrid inflation model should
satisfy in order to be in agreement with observations. First of all,
inflation should end in less than one e-fold, otherwise unacceptable
black holes would form~\cite{GBLW}. This can be written as the 
\textit{waterfall condition}~\cite{hybrid}, $\lambda m M_P^2 \gg M^3$, 
which becomes
\begin{eqnarray}
\indent {m\over M} \gg {v^2\over M_P^2}\,.
\end{eqnarray} 
Then there is the condition, known as the COBE normalization, that the
scalar amplitude should satisfy $A_S = H^2/2\pi\dot\phi \simeq
5\times10^{-5}$, which gets translated into
\begin{eqnarray}
\indent g = (n-1){M_P\over v}\sqrt{3\pi\over8}\times10^{-4}\,e^{(n-1)N/2}
\end{eqnarray} 
as well as the spectral tilt, 
\begin{eqnarray}
\indent n - 1 = {1\over\pi}{m^2\over M^2}{M_P^2\over v^2} < 0.05
\end{eqnarray} 
and finally the fact that we have not seen so far any tensor
(gravitational wave) contribution in the CMB anisotropies, $r =
A_T^2/A_S^2 < 0.3$, imposes the constraint
\begin{eqnarray}
\indent \lambda^{1/4} < 2\times10^{-3}\,{M_P\over v}\,.
\end{eqnarray} 

Taking all these conditions together, we find that a model with $v=0.1
M_P$ is probably ruled out, while one with $v=0.01\,M_P$ is perfectly
consistent with all observations, and with reasonable values of the
coupling constants, e.g. $g=4\times10^{-4}$ and $\lambda = 10^{-3}$.
However, the lower is the scale of inflation, the more difficult it is
to accommodate the amplitude of the CMB anisotropies with reasonable
values of the parameters. For a scale of inflation as low as $10^{11}$
GeV, one must significantly finetune the couplings, although there are
low scale models based on supersymmetric extensions of the standard
model which can provide a good match to observations~\cite{MSSMinf}.

In the following sections we will show how efficient is the production
of GW at reheating after hybrid inflation, using both analytical
estimates and numerical simulations to derive the amplitude of the
present day GWB. Reheating in hybrid inflation~\cite{hybrid} goes
through four well defined regimes: first, the exponential growth of
long wave modes of the Higgs field via spinodal instability, which
drives the explosive growth of all particles coupled to it, from
scalars~\cite{tachyonic} to gauge fields~\cite{CEWB} and
fermions~\cite{ester}; second, the nucleation and collision of high
density contrast and highly relativistic bubble-like structures
associated with the peaks of a Gaussian random field like the
Higgs, see paper I; third, the turbulent regime that ensues after
all these `bubbles' have collided and the energy density in all fields
cascades towards high momentum modes; finally, thermalization of all
modes when local thermal and chemical equilibrium induces
equipartition. The first three stages can be studied in detailed
lattice simulations thanks to the semi-classical character of the
process of preheating~\cite{classical}, while the last stage is
intrinsically quantum and has never been studied in the lattice.

\section{Gravitational Wave Production}

Our main purpose in this paper is to study the details of the
stochastic GWB produced during the reheating of the universe after
hybrid inflation (sections III, IV and V). However, we also study,
albeit very briefly, the analogous background from reheating in some
simple chaotic models (section VI). Thus, in this section we derive a
general formalism for extracting the GW power spectrum in any scenario
of reheating within the (flat) Friedman-Robertson-Walker (FRW)
universe. The formalism will be simplified when applied to scenarios
in which we can neglect the expansion of the universe, like in the
case of most Hybrid models.

A theory with an inflaton scalar field $\chi$ interacting 
with other Bose fields $\phi_a$, can be described by
\begin{equation}\label{lagrangian}
\indent \mathcal{L} = \frac{1}{2}\partial_\mu\chi\partial^\mu\chi +
\frac{1}{2}\partial_\mu\phi_a\partial^\mu\phi_a + \frac{R}{16\pi G} -
V(\phi,\chi)\,
\end{equation}
with $R$ the Ricci scalar. For hybrid models, we consider a generic
symmetry breaking `Higgs' field with $N_c$ real components. Thus, we
can take $\Phi^{\dagger}\Phi = {1\over2}\sum_a\phi_a^2 \equiv
|\phi|^2/2$ in (\ref{higgsPotential}), with $a$ running for the number
of Higgs' components, \textit{e.g.} $N_c = 1$ for a real scalar Higgs,
$N_c = 2$ for a complex scalar Higgs or $N_c = 4$ for a $SU(2)$ Higgs,
etc. The effective potential~(\ref{higgsPotential}) then becomes
\begin{equation}\label{higgsPotentialII}
\indent V(\phi,\chi) = \frac{\lambda}{4}\left(|\phi|^2 - 
v^2\right)^2 + g^2\chi^2|\phi|^2
+ \frac{1}{2}\mu^2\chi^2\,.
\end{equation}
For chaotic scenarios, we 
consider a massless scalar field interacting with the 
inflaton via
\begin{equation}
\indent V(\chi,\phi) = \frac{1}{2}g^2\chi^2\phi^2 + V(\chi)\,,
\end{equation}
with $V(\chi)$ the inflaton's potential. Concerning the 
simulations we show in this paper, we concentrate in the 
$N_c = 4$ case for the hybrid model and consider a potential 
$V(\chi) = {\lambda\over4}\chi^4$ for the chaotic scenario.

The classical equations of motion of the
inflaton and the other Bose fields are
\begin{eqnarray}
\label{inflatonEq}
&&\ddot\chi + 3H\dot\chi - \frac{1}{a^2}\nabla^2\chi + \frac{\partial
V}{\partial\chi} = 0 \\
\label{scalarEq}
&&\ddot\phi_a + 3H\dot\phi_a - \frac{1}{a^2}\nabla^2\phi_a +
\frac{\partial V}{\partial\phi_a} = 0
\end{eqnarray}
with $H = \dot a/a$. 

Gravitational Waves are represented here by a transverse-traceless 
(TT) gauge-invariant metric perturbation,
$h_{ij}$, on top of the flat FRW space
\begin{equation}
\indent ds^2 = -dt^2 + a^2(t)\left(\delta_{ij} + 
h_{ij}\right)dx^idx^j\,,
\end{equation}
with $a(t)$ the scale factor and the tensor perturbations verifying 
$\partial_ih_{ij} = h_{ii} = 0$. In the following, we will raise 
or low indices of the metric perturbations with the delta Kronecker 
$\delta_{ij}$, so $h_{ij} = h^{i}_{j} = h^{ij}$ and so on.
The Einstein field equations 
can be splitted into the
background $G_{\mu\nu}^{(0)} = 8\pi G\,T_{\mu\nu}^{(0)}$ and the
perturbed $\delta G_{\mu\nu} = 8\pi G\,\delta {\rm T}_{\mu\nu}$
equations. The background equations describe the evolution of the flat
FRW universe through
\begin{eqnarray}
\label{hubbleDotEq}
-\frac{\dot H}{4\pi G} &=&  \dot\chi^2 +
\frac{1}{3a^2}(\nabla\chi)^2 + \dot\phi_a^2 +
\frac{1}{3a^2}(\nabla\phi_a)^2\\
\label{hubbleEq}
\frac{3H^2}{4\pi G} &=& \dot\chi^2 +
\frac{1}{a^2}(\nabla\chi)^2 + \dot\phi_a^2 +
\frac{1}{a^2}(\nabla\phi_a)^2 + 2V(\chi,\phi)\, \nonumber \\
\end{eqnarray}
where any term in the r.h.s. of~(\ref{hubbleDotEq}) and~(\ref{hubbleEq}), 
should be understood as spatially averaged.

On the other hand, the perturbed Einstein equations describe the evolution 
of the tensor perturbations~\cite{Mukhanov} as
\begin{equation}\label{GWeq}
\indent \ddot h_{ij} + 3H\dot h_{ij} - \frac{1}{a^2}\nabla^2h_{ij} =
16\pi G\,\Pi_{ij}\,,
\end{equation}
with $\partial_i\Pi_{ij} = \Pi_{ii} = 0$. 
The source of the GW, $\Pi_{ij}$, 
contributed by both the inflaton and the other scalar fields, 
will be just the transverse-traceless part of the (spatial-spatial) 
components of the total anisotropic stress-tensor
\begin{equation}\label{ast}
\indent {\rm T}_{\mu\nu} = \frac{1}{a^2}\left[ \partial_\mu\chi
\partial_\nu\chi + \partial_\mu\phi_a\partial_\nu\phi_a + 
g_{\mu\nu}(\mathcal{L} - \left\langle p \right\rangle)\right]\,,
\end{equation}
where $\mathcal{L}(\chi,\phi_a)$ is the lagrangian~(\ref{lagrangian})
and $\left\langle p \right\rangle$ is the background homogeneous
pressure. As we will explain in the next subsection, when extracting
the TT part of~(\ref{ast}), the term proportional to $g_{\mu\nu}$ in
the r.h.s of~(\ref{ast}), will be dropped out from the GW equations
of motion. Thus, the effective source of the GW will be just
given by the TT part of the gradient terms
$\partial_\mu\chi\partial_\nu\chi +
\partial_\mu\phi_a\partial_\nu\phi_a$.

In summary, Eqs.~(\ref{inflatonEq})-(\ref{scalarEq}), together with
Eqs.~(\ref{hubbleDotEq})-(\ref{hubbleEq}), describe the coupled
dynamics of reheating in any inflationary scenario, while
Eq.~(\ref{GWeq}) describe the production of GW in each of those
scenarios. In this paper we use lattice simulations to study the
generation of GW during reheating after inflation. Specific details on
this are given in section~IV, but let us just mention here that our
approach is to solve the evolution of the gravitational waves
simultaneously with the dynamics of the scalar fields, in a
discretized lattice with periodic boundary conditions. We assume
initial quantum fluctuations for all fields and only a zero mode for
the inflaton.  Moreover, we also included the GW backreaction on the
scalar fields' evolution via the gradient terms,
$h^{ij}\nabla_i\chi\nabla_j\chi + h^{ij}\nabla_i\phi_a\nabla_j\phi_a$
and confirmed that, for all practical purposes, these are negligible
throughout GW production.

\subsection{The Transverse-Traceless Gauge}

A generic (spatial-spatial) metric perturbation $\delta h_{ij}$ has six 
independent degrees of freedom, whose contributions can be split into 
scalar, vector and tensor metric perturbations~\cite{Mukhanov}
\begin{equation}
\indent
\delta h_{ij} = \psi\,\delta_{ij} + E_{,ij} + F_{(i,j)} + h_{ij}\,.
\end{equation}
with $\partial_iF_i = 0$ and $\partial_ih_{ij} = h_{ii} = 0$.  By
choosing a transverse-traceless stress-tensor source $\Pi_{ij}$, we
can eliminate all the degrees of freedom (d.o.f.) but the pure TT
part, $h_{ij}$, which represent the only physical d.o.f which
propagate and carry energy out of the source (GW). If we had chosen
only a traceless but non-transverse stress source, we could have
eliminated the scalar d.o.f.  $\psi$ and absorbed $E$ into the scalar
field perturbation, but we would still be left with a vector field
$F_i$ also sourced by the (traceless but non-transverse) anisotropic
stress tensor, thus giving rise to a vorticity divergence-less field
$V_i$. However, since the initial conditions are those of a scalar
Gaussian random field (see section IV), even in that case of a
non-transverse but traceless stress source, the mean vorticity of the
subsequent matter distribution, averaged over a sufficiently large
volume, should be zero (although locally we do have vortices of the
Higgs field, see Refs.~\cite{CEWB,magnetic}), since vortices with
opposite chirality cancel eachother. This means that in such a case,
although $\partial^i\Pi_{ij} \neq 0$, and thus $\partial^i\delta
h_{ij} \neq 0$, we could still recover the TT component when averaging
over the whole realization.

For practical purposes, we will consider from the beggining the TT
part of the anisotropic stress-tensor, ensuring this way that we only
source the physical d.o.f. that represent GW. The equations of motion
of the TT metric perturbations are then given by Eq.~(\ref{GWeq}).
Note that for a non-transverse source the equations would have been
much more complicated, so the advantage of using the TT part from the
beginning is clear. The disadvantage arises because obtaining the TT
part of a tensor in configuration space is very demanding in
computational time. However, as we explain next, we will use a method by
which we can circunvent this issue. 

Let us switch to Fourier space. Using the convention
\begin{equation}
\indent f(\mathbf{k}) = \frac{1}{(2\pi)^{3/2}}\int d^3\mathbf{x}\,
e^{+i\mathbf{kx}}f(\mathbf{x})\,,
\end{equation} 
the GW equations~(\ref{GWeq}) in Fourier space read
\begin{equation}\label{GWeqFourier}
\indent \ddot h_{ij}(t,\mathbf{k}) + 3H\dot h_{ij}(t,\mathbf{k}) + 
\frac{k^2}{a^2}h_{ij}(t,\mathbf{k}) = 16\pi G\,\Pi_{ij}(t,\mathbf{k})\,,
\end{equation}
where $k = |\mathbf{k}|$. Assuming no GW at the beginnig of reheating
(i.e.  the end of inflation $t_e$), the initial conditions are
$h_{ij}(t_e) = \dot h_{ij}(t_e) = 0$, so the solution to
Eq.~(\ref{GWeqFourier}) for $t > t_e$ will be just given by a causal
convolution with an appropriate green function $G(t,t')$,
\begin{eqnarray}\label{GWsol}
\indent h_{ij}(t,\mathbf{k}) = 16\pi G \int_{t_e}^{t}dt'\,G(t,t')
\Pi_{ij}(t',\mathbf{k})
\end{eqnarray} 
Therefore, all we need to know for computing the GW is the TT part of
the stress-tensor, $\Pi_{ij}$, and the Green function $G(t',t)$.
However, as we will demonstrate shortly, we have used a numerical
method by which we don't even need to know the actual form of
$G(t',t)$.  To see this, let us extract the TT part of the total
stress-tensor. Given the symmetric anisotropic stress-tensor
T$_{\mu\nu}$~(\ref{ast}), we can easily obtain the TT part of its
spatial components in momentum space, $\Pi_{ij}(\mathbf{k})$. Using
the spatial projection operators $P_{ij} = \delta_{ij} - \hat k_i \hat
k_j$, with $\hat k_i = k_i/k$, then~\cite{Carroll}
\begin{eqnarray}
\label{TTgauge}
\indent \Pi_{ij}(\mathbf{k}) = \Lambda_{ij,lm}(\mathbf{\hat k})
{\rm T}_{lm}(\mathbf{k})\,,
\end{eqnarray} 
where
\begin{eqnarray}\label{projector}
\indent \Lambda_{ij,lm}(\mathbf{\hat k}) \equiv \Big(P_{il}(\mathbf{\hat k})
P_{jm}(\mathbf{\hat k}) - \half P_{ij}(\mathbf{\hat k})
P_{lm}(\mathbf{\hat k})\Big)\,.
\end{eqnarray}
Thus, one can easily see that, at any time $t$, $k_i\Pi_{ij}(\mathbf{\hat
k},t) = \Pi_{i}^{i}(\mathbf{\hat k},t) = 0$, as required, thanks to
the identities $P_{ij}\hat k_j = 0$ and $P_{ij}P_{jm} = P_{im}$.

Note that the TT tensor, $\Pi_{ij}$, is just a linear combination of
the components of non-traceless nor-transverse tensor
T$_{ij}$~(\ref{ast}), while the solution~(\ref{GWsol}) is just linear
in $\Pi_{ij}$. Therefore, we can write the TT tensor perturbations 
(i.e. the gravitational waves) as
\begin{eqnarray}
\label{TTsol}
\indent h_{ij}(t,\mathbf{k}) = \Lambda_{ij,lm}(\mathbf{\hat k}) 
 u_{ij}(t,\mathbf{k})\,,
\end{eqnarray} 
with $u_{ij}(t,\mathbf{k})$ the Fourier transform of the solution 
of the following equation
\begin{eqnarray}\label{GWfakeEq}
\indent \ddot u_{ij} + 3H\dot u_{ij} - 
\frac{1}{a^2}\nabla^2 u_{ij} = 16\pi G\,{\rm T}_{ij}
\end{eqnarray}
This Eq.~(\ref{GWfakeEq}) is nothing but Eq.~(\ref{GWeq}), sourced
with the complete T$_{ij}$~(\ref{ast}), instead of with its TT part,
$\Pi_{ij}$.  Of course, Eq.~(\ref{GWfakeEq}) contains unphysical
(gauge) d.o.f.; however, in order to obtain the real physical TT
d.o.f., $h_{ij}$, we can evolve Eq.(\ref{GWfakeEq}) in configuration
space, Fourier transform its solution and apply the
projector~(\ref{projector}) as in ~(\ref{TTsol}). This way we can
obtain in momentum space, at any moment of the evolution, the physical
TT d.o.f. that represent GW, $h_{ij}$. Whenever needed, we can Fourier
transform back to configuration space and obtain the spatial
distribution of the gravitational waves.

Moreover, since the second term of the r.h.s of the total
stress-tensor T$_{ij}$ is proportional to $g_{ij} = \delta_{ij} +
h_{ij}$, see~(\ref{ast}), when aplying the TT
projector~(\ref{projector}), the part with the $\delta_{ij}$ just
drops out, simply because it is a pure trace, while the other part 
contributes with a term $-(\mathcal{L} - \left\langle p
\right\rangle)h_{ij}$ in the l.h.s of Eq.(\ref{GWeqFourier}). 
However, $(\mathcal{L} - \left\langle p \right\rangle)$ is of the same 
order as the metric perturbation $\sim {\cal O}(h)$, 
so this extra term is second order in the gravitational coupling 
and it can be neglected in the GW Eqs.~(\ref{GWeqFourier}).
This way, the effective source in Eq.~(\ref{GWfakeEq}) is just the 
gradient terms of both the inflaton and the other scalar fields,
\begin{eqnarray}\label{GWfakeSource}
\indent {\rm T}_{ij} = \frac{1}{a^2}\left(\nabla_i\chi\nabla_j\chi + 
\nabla_i\phi_a\nabla_j\phi_a\right)
\end{eqnarray} 
Therefore, the effective source of the physical GW, will be just the
TT part of~(\ref{GWfakeSource}), as we had already mentioned before.

Alternatively, one could evolve the equation of the TT tensor
perturbation in configuration space, Eq.~(\ref{GWeq}), with the source
given by
\begin{equation}\label{GWtrueSource}
\Pi_{ij}({\bf x},t) = {1\over(2\pi)^{3/2}}\int d^3{\bf k}\,
e^{-i{\bf kx}}\Lambda_{ij,lm}({\bf \hat k}){\rm T}_{lm}({\bf k},t)\,, 
\end{equation}
such that $\partial_i\Pi_{ij}({\bf x},t) = \Pi_{ii}({\bf x},t) = 0$ at
any time.  So, at each time step of the evolution of the fields, one
would have first to compute (the gradient part of)
T$_{lm}$~(\ref{GWfakeSource}) in configuration space, then Fourier
transform it to momentum space, substitute in Eq.~(\ref{GWtrueSource})
and perform the integral. However, proceeding as we suggested above,
there is no need to perform the integral,
nor Fourier transform the fields at each time step, but rather only at
those times at which we want to measure the GW spectrum. The viability
of our method relies in the following observation. To compute the GW
we could, first of all, project the TT part of the
source~(\ref{GWtrueSource}), and second, solve Eq.~(\ref{GWeq}).
However, we achieve the same result if we commute these two operations
such that, first, we solve Eq.~(\ref{GWfakeEq}), and second, we apply
the TT projector to the solution~(\ref{TTsol}). We have found this
\textit{commuting procedure} very useful, since we are able to extract
the spectra or the spatial distribution of the GW at any desired time,
saving a great amount of computing time. Most importantly, with this
procedure we can take into account backreaction simultaneously with
the fields evolution.

In summary, for solving the dynamics of reheating of a particular
inflationary model, we evolve Eqs.~(\ref{inflatonEq})-(\ref{scalarEq})
in the lattice, together with
Eqs.~(\ref{hubbleDotEq})-(\ref{hubbleEq}), while for the GWs 
we solve Eq.~(\ref{GWfakeEq}). Then, only
when required, we Fourier transform the solution of
Eq.~(\ref{GWfakeEq}) and then apply~(\ref{TTsol}) in order to recover
the physical transverse-traceless d.o.f representing the GW. From
there, one can easily build the GW spectra or take a snapshot of
spatial distribution of the gravitational waves.

\subsection{The energy density in GW}

The energy-momentum tensor of the GW is given by~\cite{Carroll}
\begin{equation}\label{tmunu}
\indent t_{\mu\nu} = {1\over 32\pi G}\,\left\langle\partial_\mu
h_{ij}\, \partial_\nu h^{ij}\right\rangle_{\rm V}\,,
\end{equation}
where $h_{ij}$ are the TT tensor perturbations solution of
Eq.(\ref{GWeq}). The expectation value
$\left\langle...\right\rangle_{\rm V}$ is taken over a region of
sufficiently large volume $V=L^3$ to encompass enough physical
curvature to have a gauge-invariant measure of the GW energy-momentum
tensor.

The GW energy density will be just 
$\rho_{_{\rm GW}} = t_{00}$, so
\begin{eqnarray}\label{rho}
\indent \rho_{GW} &=& \frac{1}{32\pi G}\frac{1}{L^3}\int d^3\mathbf{x}\,
\dot h_{ij}(t,\mathbf{x})\dot h_{ij}(t,\mathbf{x}) \nonumber \\ 
&=&  \frac{1}{32\pi G}\frac{1}{L^3}\int d^3\mathbf{k}\,
\dot h_{ij}(t,\mathbf{k})\dot h_{ij}^*(t,\mathbf{k})\,,
\end{eqnarray} 
where in the last step we Fourier transformed each 
$\dot h_{ij}$ and used the integral definition of the Dirac delta 
$(2\pi)^3\delta^{(3)}(\mathbf{k}) = \int d^3\mathbf{x}\ e^{-i\mathbf{kx}}$.

We can always write the scalar product in (\ref{rho}) in terms of the 
(Fourier transformed) solution $u_{lm}$ of the Eq.(\ref{GWfakeEq}), 
by just using the
spatial projectors (\ref{projector})
\begin{equation}
\indent \dot h_{ij}\dot h_{ij} = \Lambda_{ij,lm}\Lambda_{ij,rs} 
\dot u_{lm} \dot u_{rs} = 
\Lambda_{lm,rs} \dot u_{lm} \dot u_{rs}\,,
\end{equation}
where we have used the fact that $\Lambda_{ij,lm}\Lambda_{lm,rs} =
\Lambda_{ij,rs}$. This way, we can express the GW energy density as
\begin{eqnarray}\label{rhoTotal}
&&\rho_{GW} = \frac{1}{32\pi G L^3}\times \nonumber \\ 
&&\hspace{1.2cm} \int k^2dk \int d\Omega\,\Lambda_{ij,lm}(\mathbf{\hat k})
\dot  u_{ij}(t,\mathbf{k})\dot  u_{lm}^*(t,\mathbf{k}).
\end{eqnarray}
From here, we can also compute the power spectrum per logarithmic frequency
interval in GW, normalized to the critical density $\rho_c$, as
\begin{equation}
\indent \Omega_{_{\rm GW}}=\int {df\over f}\,\Omega_{_{\rm GW}}(f)\,, 
\end{equation}
where
\begin{eqnarray}\label{OmegaFraction}
\Omega_{gw}(k) \equiv \frac{1}{\rho_c}\frac{d\rho_{gw}}{d\,{\rm log}k}\hspace{5.2cm} \nonumber \\
= \frac{k^3}{32\pi GL^3\rho_c} \int d\Omega\,\Lambda_{ij,lm}(\mathbf{\hat k})
\dot  u_{ij}(t,\mathbf{k})\dot  u_{lm}^*(t,\mathbf{k})\hspace{.4cm} 
\end{eqnarray}
We have checked explicitely in the simulations that the argument 
of the angular integral of~(\ref{OmegaFraction}) is independent 
of the directions in {\bf k}-space. Thus, whenever we plot the GW 
spectrum of any model, we will be showing the amplitude of the 
spectrum (per each mode $k$) as obtained after avaraging over all 
the the directions in momentum space,
\begin{eqnarray}
\indent\Omega_{gw}(k) = \frac{k^3}{8GL^3\rho_c} \left\langle \Lambda_{ij,lm}(\mathbf{\hat k})
\dot  u_{ij}(t,\mathbf{k})\dot  u_{lm}^*(t,\mathbf{k})\right\rangle_{4\pi} 
\end{eqnarray} 
where $\left\langle f \right\rangle_{4\pi} \equiv {1\over 4\pi}\int f{\rm d}\Omega$. 

Finally, we must address the fact that the frequency range, for a GWB
produced in the early universe, will be redshifted today. We should
calculate the characteristic physical wavenumber of the present GW
spectrum, which is redshifted from any time $t$ during GW
production. This is a key point, since a relatively long period of
turbulence will develop after preheating, which could change the
amplitude of the GWB and shift the frequency at which the spectra
peaks. So let us distinguish four characteristic times: the end of
inflation, $t_e$; the time $t_*$ when GW production stops;
the time $t_{r}$ when the universe finally reheats and enters into the
radiation era; and today, $t_0$.\footnote{Note, however, that after
thermalization there is still a small production of GW from the
thermal plasma, but this can be ignored for all practical purposes.}
Thus, today's frequency $f_0$ is related to the physical wavenumber
$k_t$ at any time $t$ of GW production, via $f_0 =
(a_t/a_0)(k_t/2\pi)$, with $a_0$ and $a_t$, the scale factor today and
at the time $t$, respectively. Thermal equilibrium was established at
some temperature $T_r$, at time $t_r \geq t$. The Hubble rate at that
time was $M_P^2H_r^2 = (8\pi/3)\rho_r$, with $\rho_r =
g_r\pi^2T_r^4/30$ the relativistic energy density and $g_r$ the
effective number of relativistic degrees of freedom at temperature
$T_r$. Since then, the scale factor has increased as $a_r/a_0 =
(g_{0,s}/g_{r,s})^{1/3}(T_0/T_{r})$, with $g_{i,s}$ the effective
entropic degrees of freedom at time $t_i$, and $T_0$
today's CMB temperature. Putting all together,
\begin{eqnarray}
\label{redshift}
\indent f_0 = \left(\frac{8\pi^3g_{r}}{90}\right)^{1\over4}
\left(\frac{g_{0,s}}{g_{{r},s}}\right)^{1\over3}\frac{T_0}
{\sqrt{H_{r}M_p}}\left(\frac{a_e}{a_r}\right)\frac{k}{2\pi}\,,
\end{eqnarray}
where we have used the fact that the physical wave number $k_t$ at 
any time $t$ during GW production, is related to the comoving wavenumber $k$
through $k_t = (a_e/a_t)k$ with the normalization $a_e \equiv 1$.

From now on, we will be concerned with hybrid inflation, leaving
chaotic inflation for section VI. Within the hybrid scenario, we will
analyse the dependence of the shape and amplitude of the produced GWB
on the scale of hybrid inflation, and more specifically on the
\textit{v.e.v.} of the Higgs field triggering the end of inflation.
The initial conditions are carefully treated following the
prescription adopted in paper I, as explained in section IV. Given the
natural frequency at hand in hybrid models, $m = \sqrt{\lambda}v$,
whose inverse $m^{-1}$ sets the characteristic time scale during the
first stages of reheating, it happens that as long as $v\,\ll\,M_p$,
the Hubble rate $H\,\sim\,\sqrt{\lambda}(v^2/M_p)$ is much smaller
than such a frequency, $H\,\ll\,m$. Indeed, all the initial vacuum
energy $\rho_0$ gets typically converted into radiation in less than a
Hubble time, in just a few $m^{-1}$ time steps. Therefore, we should
be able to ignore the dilution due to the expansion of the universe
during the production of GW, at least during the first stages of
reheating. However, as we will see, the turbulent behaviour developed
after those first stages, could last for much longer than an e-fold,
in which case we will have to take into account the expansion of the
universe. Our approach will be first to ignore the expansion of the
Universe and later see how we can account for corrections if
needed. Thus, we set the scale factor $a = 1$ and the Hubble rate $H =
0$ and $\dot H = 0$. As we will see later, our approach of neglecting 
the expansion for the time of GW production, will be completely justified
\textit{a~posteriori}.

The coupled evolution equations that
we have to solve numerically in a lattice for the hybrid model are
\begin{eqnarray}\label{GWeqHybrid}
&&\ddot u_{ij} - \nabla^2 u_{ij} = 16\pi G\,{\rm T}_{ij} \\
&&\ddot\chi - \nabla^2\chi + \left(g^2|\phi|^2 +
\mu^2\right)\chi = 0 \\ &&\ddot\phi_a - \nabla^2\phi_a +
\left(g^2\chi^2 + \lambda|\phi|^2 - m^2\right)\phi_a = 0
\end{eqnarray}
with T$_{ij}$ given by Eq.(\ref{GWfakeSource}) with the scale factor
$a = 1$. We have explicitly checked in our computer simulations that
the backreaction of the gravity waves into the dynamics of both the
inflaton and the Higgs fields is negligible and can be safely ignored.
We thus omit the backreaction terms in the above equations.

\begin{figure}[t]
\begin{center}
\includegraphics[width=5.5cm,height=8.5cm,angle=-90]{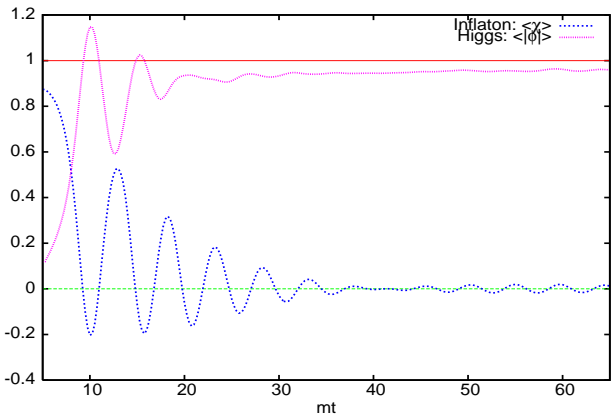}
\end{center}
\vspace*{-5mm}
\caption{Time evolution of the mean field values of the Higgs and the
Inflaton, the former normalized to its \textit{v.e.v.}, the latter
normalized to its critical value $\chi_{0} = \rm{m/g}$. This is just a
specific realization with $N = 128$, $p_{\rm min} = 0.1m$, $a =
0.48m^{-1}$, $v = 10^{-3}M_p$ and $g^2 = 2\lambda = 0.25$.}
\label{fig1}
\vspace*{-3mm}
\end{figure}

We evaluate during the evolution of the system the mean field values,
as well as the different energy components. As shown in
Fig.~\ref{fig1}, the Higgs field grows towards the true vacuum and
the inflaton moves towards the minimum of its potential and oscillates
around it. We have checked that the sum of the averaged gradient,
kinetic and potential energies (contributed by both the inflaton and
the Higgs), remains constant during reheating, as expected, since the
expansion of the universe is irrelevant in this model. We have also
checked that the time evolution of the different energy components is
the same for different lattices, i.e. changing the number of points
$N$ of the lattice, of the minimum momentum $p_{\rm min} = 2\pi/L$ or
of the lattice spacing $a = L/N$, with $L$ the lattice size, as long
as the product $ma < 0.5$; for a detailed discussion of lattice scales
see paper I.  Looking at the time evolution of the Higgs'
\textit{v.e.v.}  in  Fig.~\ref{fig1}, three stages can be
distinguished. First, an exponential growth of the \textit{v.e.v.}
towards the true vacuum. This is driven by the tachyonic instability
of the long-wave modes of the Higgs field, that makes the spatial
distribution of this field to form lumps and bubble-like
structures~\cite{tachyonic,symmbreak}. Second, the Higgs field
oscillates around the true vacuum, as the Higgs' bubbles collide and
scatter off eachother. Third, a period of turbulence is reached,
during which the inflaton oscillates around its minimum and the Higgs
sits in the true vacuum. For a detailed description of the dynamics of
these fields see Ref.~\cite{symmbreak}. Here we will be only concerned
with the details of the gravitational wave production.

\begin{figure}[t]
\begin{center}
\includegraphics[width=5.5cm,height=8.5cm,angle=-90]{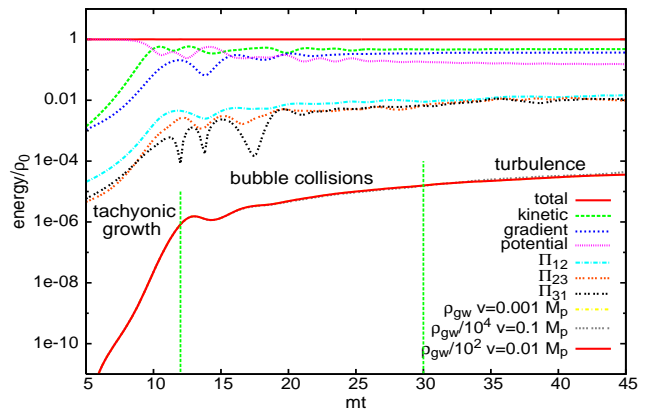}
\end{center}
\vspace*{-5mm}
\caption{The time evolution of the different types of energy (kinetic,
gradient, potential, anisotropic components and gravitational waves
for different lattices), normalized to the initial vacuum energy,
after hybrid inflation, for a model with $v=10^{-3}\,M_P$. One can
clearly distinguish here three stages: tachyonic growth, bubble
collisions and turbulence. }
\label{fig3}
\vspace*{-3mm}
\end{figure}

The initial energy density at the end of hybrid inflation is given by
$\rho_0 = m^2v^2/4$, with $m^2 = \lambda v^2$, so the fractional
energy density in gravitational waves is
\begin{equation}
\indent {\rho_{_{\rm GW}}\over\rho_0} = {4t_{00}\over v^2\,m^2} =
\frac{1}{8\pi\,G v^2 m^2}\left\langle\dot h_{ij}\dot
h^{ij}\right\rangle_{\rm V}\,,
\end{equation}
where $\left\langle\dot h_{ij}\dot h^{ij}\right\rangle_V$, defined as a 
volume average like 
${1\over {\rm V}}\int d^3{\bf x}\dot h_{ij}\dot h^{ij}$, is extracted from 
the simulations as
\begin{eqnarray}
&&\left\langle\dot h_{ij}\dot
h^{ij}\right\rangle_{\rm V} = \nonumber \\ && = 
\dfrac{4\pi}{V}\int d{\rm log}k\,k^3\left\langle \Lambda_{ij,lm}(\mathbf{\hat k})
\dot  u_{ij}(t,\mathbf{k})\dot  u_{lm}^*(t,\mathbf{k})
\right\rangle_{4\pi} 
\end{eqnarray}
where $ u_{ij}(t,\mathbf{k})$ is the Fourier transform of the solution 
of Eq.~(\ref{GWeqHybrid}). 
Then, we can compute the corresponding
density parameter today (with $\Omega_{\rm
rad}\,h^2\simeq3.5\times10^{-5}$)
\begin{eqnarray}
&& \Omega_{_{\rm GW}}\,h^2 = {\Omega_{\rm rad}\,h^2\over 2
G\,v^2\,m^2\,V}\,\times \nonumber \\ &&\hspace{.7cm} 
\int d{\rm log}k\, k^3\left\langle \Lambda_{ij,lm}(\mathbf{\hat k})
\dot  u_{ij}(t,\mathbf{k})\dot  u_{lm}^*(t,\mathbf{k})
\right\rangle_{4\pi} 
\end{eqnarray}
which has assumed that all the
vacuum energy $\rho_0$ gets converted into radiation, an approximation
which is always valid in generic hybrid inflation models with $v\ll
M_P$, and thus $H\ll m=\sqrt\lambda\,v$.

We have shown in Fig.~\ref{fig3} the evolution in time of the fraction
of energy density in GW. The first (tachyonic) stage is clearly
visible, with a (logarithmic) slope twice that of the anisotropic
tensor $\Pi_{ij}$. Then there is a small plateau corresponding to the
production of GW from bubble collisions; and finally there is the
slow growth due to turbulence. In the next section we will describe
in detail the most significant features appearing at each stage.

Note that in the case that $H\ll m$, the maximal production of GW
occurs in less than a Hubble time, soon after symmetry breaking, while
turbulence lasts several decades in time units of $m^{-1}$. Therefore,
we can safely ignore the dilution due to the Hubble expansion, up to
times much greater than those of the tachyonic instability. Eventually
the universe reheats and the energy in gravitational waves redshifts
like radiation thereafter.

\begin{figure}[t]
\vspace{-4mm}
\begin{center}
\includegraphics[width=5.5cm,height=8.7cm,angle=-90]{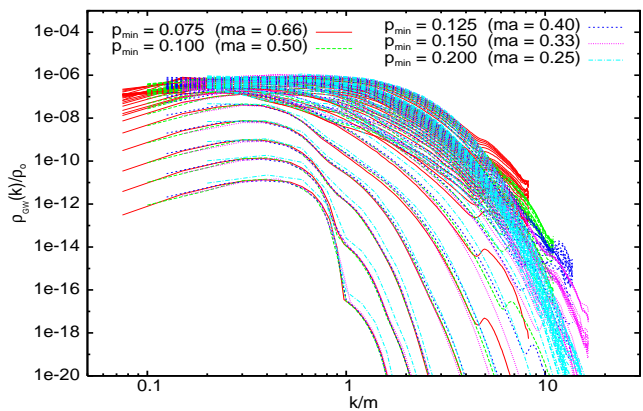}
\end{center}
\vspace*{-5mm}
\caption{We show here the comparison between the power spectrum of
gravitational waves obtained with increasing lattice resolution, to
prove the robustness of our method. The different realizations are
characterized by the 
the minimum 
lattice momentum (p$_{\rm min}$) and the lattice spacing (ma). The
growth is shown in steps of $m\Delta t = 1$ up to $mt = 30$, and 
then in and $m\Delta t = 5$ steps up to $mt = 60$.}
\label{fig4}
\vspace*{-3mm}
\end{figure}

To compute the power spectrum per logarithmic frequency
interval in GW, $\Omega_{gw}(f)$, we just have to use~(\ref{OmegaFraction}).
Moreover, since gravitational waves below Planck scale remain decoupled 
from the plasma immediately after
production, we can evaluate the power spectrum today from that
obtained at reheating by converting the wavenumber $k$ into
frequency $f$. Simply using Eq.~(\ref{redshift}), with 
$g_{{r},s}/g_{0,s}\sim 100$, $g_{{r},s}\,\sim\,g_{r}$ and 
$a_e \sim a_*$, then
\begin{equation}
\indent f=6\times10^{10}\,{\rm Hz}\,{k\over\sqrt{H\,M_p}}
=5\times10^{10}\,{\rm Hz}\,{k\over m}\,\lambda^{1/4}\,.
\end{equation}
We show in Fig.~\ref{fig4} the power spectrum of gravitational waves
as a function of (comoving) wavenumber $k/m$. We have used different
lattices in order to have lattice artifacts under control, specially
at late times and high wavenumbers. We made sure by the choice of
lattice size and spacing (i.e. $k_{\rm min}$ and $k_{\rm max}$) that
all relevant scales fitted within the simulation. Note, however, that
the lower bumps are lattice artifacts, due to the physical cutoff
imposed at the initial condition, that rapidly disappear with time. We
have also checked that the power spectrum of the scalar fields follows
turbulent scaling after $mt \sim {\cal O}(100)$, and we can thus
estimate the subsequent evolution of the energy density distributions
beyond our simulations.

\section{Lattice simulations}

The problem of determining the time evolution of a quantum field
theory is an outstandingly difficult problem. In some cases only a few
degrees of freedom are relevant or else perturbative techniques are
applicable. However, in our particular case, our interests are focused
on processes which are necessarily non-linear and non-perturbative and
involve many degrees of freedom. The presence of gravitational fields
just contributes with more degrees of freedom, but does not complicate 
matters significantly.

The lattice formulation allows a first principles approach to
non-perturbative quantum field theory. The existing powerful lattice
field theory numerical methods rest on the path integral formulation
in Euclidean space and the existence of a probability measure in field
space~\cite{Wilson}. However, the problem we are interested in is a
dynamical process far from equilibium, and the corresponding Minkowski
path integral formulation is neither mathematically well founded nor
appropriate for numerical studies. There are a series of alternative
non-perturbative methods which different research groups have used to
obtain physical results in situations similar to ours. These include
Hartree's approximations~\cite{Cooper} to go beyond perturbation
theory or large N techniques~\cite{Vega,Baacke}. It is clear that it
is desirable to look at this and similar problems with all available
tools. In the present paper we will use an alternative approximation
to deal with the problem: the classical approximation.  It consists of
replacing the quantum evolution of the system by its classical
evolution, for which there are feasible numerical methods
available. The quantum nature of the problem remains in the stochastic
character of the initial conditions. This approximation has been used
with great success by several groups in the
past~\cite{classical,tachyonic}. The advantage of the method is that
it is fully non-linear and non-perturbative, allows the use of gauge
fields~\cite{CEWB,magnetic} and gives access to the quantities we are
interested in.
 
The validity of the approximation depends on the loss of quantum
coherence in the evolution of the system. In previous papers we
studied this problem both in the absence of and with gauge
fields~\cite{symmbreak,CEWB,magnetic}. We started the evolution of the
system at the critical time $t_c$, corresponding to the end of
inflation $t_e$, at which the effective mass of the Higgs vanishes,
putting all the modes in its (free field) ground states. If the
couplings are small, since the quantum fluctuations of the value of
the fields are not too large, the non-linear terms in the Hamiltonian
of the system can be neglected. Then the quantum evolution is Gaussian
and can be studied exactly. The Hamiltonian has nonetheless a
time-dependence coming through the time-dependence of the inflaton
homogeneous mode. This time dependence can always be taken to be
linear for a sufficiently short time interval after the critical
time. As a consequence, the dynamics of the eigenmodes during this
initial phase differs significantly from mode to mode. Most of them
have a characteristic harmonic oscillator behaviour with a frequency
depending on the mode in question.  In the case of the Higgs field,
the long-wave modes become tachyonic.  By looking at expectation
values of products of these fields at different times, one realises
that after a while these modes behave and evolve like classical modes
of an exponentially growing size. The process is very fast and
therefore the remaining harmonic modes can be considered to have
remained in their initial ground state.
 
The fast growth in size of the Higgs field expectation value boosts
the importance of non-linear terms and eventually drives the system
into a state where the non-linear dynamics, including the
back-reaction to the inflaton field, are crucial. For the whole
approximation to be useful this must happen at a later time than the
one in which the low-frequency Higgs modes begin evolving as classical
fields. In paper I we showed this to be the case.  Actually, there is
a time interval in which classical behaviour is already dominant while
non-linearities are still small. We tested that our results, in the
absence of gauge fields, were insensitive to the matching time,
provided it lies within this window.
 
The whole idea can then be summarised as follows: the tachyonic
quantum dynamics of the low momentum Higgs modes drive them into
classical field behaviour and large occupation numbers before the
non-linearities and back-reaction begin to play a role. It is the
subsequent non-linear classical behaviour of the field that induces
the growth of classical inflaton and gravitational field components
also at low frequencies.  It is obvious that the quantum nature of the
problem is still crucial if one studies the behaviour of high momentum
modes which have low occupation numbers. 
 
In the present paper we apply the same idea in the presence of
(gravitational wave) tensor fields. The initial quantum evolution of
tensor fields is also relatively slow, since there are no tachyonic
modes. Therefore, it is assumed not to affect substantially the
initial conditions of the classical system.

\subsection{Initial conditions}

Our approach to the dynamics of the system is to assume that the
leading effects under study can be well-described by the classical
evolution of the system. The justification of this point, as explained
in the previous section, rests upon the fast quantum evolution of the
long wavelength components of the Higgs field during the initial
stages after the critical point. All the other degrees of freedom will
evolve slowly from their initial quantum vacuum state. For the Higgs
field, the leading behaviour is the exponential growth of those modes
having negative effective mass-squared. The quantum evolution of such
modes drives the system into a quasi-classical regime. It is essential
that this regime is reached before the non-linearities couple all
degrees of freedom to each other and questions like back-reaction
start to affect the results. It is then assumed that it is the
essentially classical dynamics of that field what matters, and that
all the long-wavelength components of the inflaton and the gauge
fields produced by the interaction with the Higgs field behave also as
classical fields. Of course, this can hardly be the case for shorter
wavelengths which stay in a quantum state with low occupation numbers.
However, as we can see in Fig.~\ref{fig4}, for the range of times
studied in this paper, the effect of shorter wavelengths is relatively
small, and the spectrum of modes remains always dominated by
long-wavelengths. 

The full non-linear evolution of the system can then be studied using
lattice techniques. Our approach is to discretize the classical
equations of motion of all fields in both space and time. The
time-like lattice spacing $a_t$ must be smaller than the spatial one
$a_s$ for the stability of the discretized equations. In addition to
the ultraviolet cut-off one must introduce an infrared cut-off by
putting the system in a box with periodic boundary conditions.  We
have studied $64^3$, $128^3$ and $256^3$ lattices.  Computer memory
and CPU resources limit us from reaching much bigger lattices.
Nonetheless, in the spirit of paper I, there are a number of
checks one can perform to ensure that our results are physical and are
not biased, within errors, by the approximations introduced, see 
Fig.~\ref{fig4}. Our problem has several physical scales which control
different time-regimes and observables. Thus, it is not always an easy
matter to place these scales in the window defined by our ultraviolet
and infrared cut-offs.  For example, in addition to the Higgs and
inflaton mass there is a scale $M$ associated to the inflaton velocity
which is particularly relevant in determining the bubble sizes and
collisions.  Then, when we want to study a stage of the evolution in
particular, we make the selection of the volume and cutoff most
suitable.

Since in this paper we are more interested in understanding the
phenomenon of GW production, rather than concentrating in a particular
model, our attitude has been to modify the parameters of the model in
order to sit in a region where our results are insensitive to the
cut-offs.  This is no doubt a necessary first step to determine the
requirements and viability of the study of any particular model. In
particular, we have thouroughly studied a model with
$g^2=2\lambda=1/4$, but we have checked that other values of the
parameters do not change our results significantly.

The initial conditions of the fields follow the prescription from
paper I. The Higgs modes $\phi_k$ are solutions of the coupled
evolution equations, which can be rewritten as
$\phi''_k+(k^2-\tau)\phi_k=0$, with $\tau=M(t-t_c)$ and
$M=(2V)^{1/3}m$. The time-dependent Higgs mass follows from the
initial inflaton field homogeneous component, $\chi_0(t_i) =
\chi_c(1-Vm(t_i-t_c))$ and $\dot\chi_0(t_i)=-\chi_cVm$. The Higgs
modes with $k/M > \sqrt{\tau_i}$ are set to zero, while the rest are
determined by a Gaussian random field of zero mean distributed
according to the Rayleigh distribution
\begin{equation}
\indent 
P(|\phi_k|)d|\phi_k|d\theta_k = \exp\left(-{|\phi_k|^2\over\sigma_k^2}
\right)\,{d|\phi_k|^2\over\sigma_k^2}\,{d\theta_k\over2\pi}\,,
\end{equation}
with a uniform random phase $\theta_k\in[0,2\pi]$ and dispersion given 
by $\sigma_k^2 \equiv |f_k|^2 = P(k,\tau_i)/k^3$, where $P(k,\tau_i)$ is the power 
spectrum of the initial Higgs quantum fluctuations in the background 
of the homogeneous inflaton, computed in the linear approximation.
In the region of low momentum modes it is well approximated by
\begin{equation}\label{Papp}
\indent 
2kP_{\rm app}(k,\tau_i) = k^3\left(1+A(\tau_i)\,k^2\,e^{-B(\tau_i)\,k^2}\right)\,,
\end{equation}
where $A(\tau_i)$ and $B(\tau_i)$ are parameters extracted from a fit
of this form to the exact power spectrum given in paper~I. In the
classical limit, the conjugate momentum $\dot\phi_k(\tau)$ is uniquely
determined through $\dot\phi_k(\tau) = F(k,\tau)\phi_k(\tau)$, where
$F(k,\tau) = {\rm Im}(if_k(\tau)\dot f_k^*(\tau))/|f_k(\tau)|^2$, see
paper I. In the region of low momenta, $F(k,\tau_i)$ can be well
aproximated by
\begin{eqnarray}
\indent F(k,\tau_i) = \frac{2kC(\tau_i)e^{-D(\tau_i)k^2}}{[1+A(\tau_i)e^{-B(\tau_i)k^2}]}.
\end{eqnarray}
where $A(\tau_i)$ and $B(\tau_i)$ are the previous coefficients for the amplitude of the field fluctuations, while $C(\tau_i)$ and $D(\tau_i)$ are new coefficients obtained fitting the exact expression of $F(k,\tau_i)$.

\begin{figure}[t]
\begin{center}
\includegraphics[width=5.5cm,height=8.5cm,angle=-90]{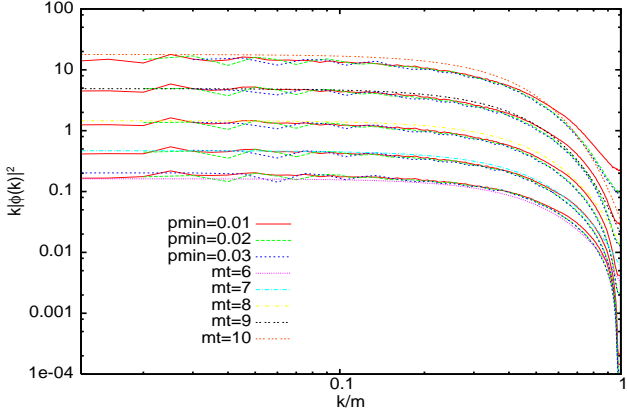}
\end{center}
\vspace*{-5mm}
\caption{The tachyonic growth of the Higgs' spectrum, from $mt=5$ to
$mt=10$. We compare simulations of different sizes ($p_{\rm min} = 
0.01 - 0.03$) and $N=256$, with the anaylitical expressions.} 
\label{HigSpec}
\vspace*{-3mm}
\end{figure}

The rest of the fields (the inflaton non-zero modes and the
gravitational waves), are supposed to start from the vacuum, and
therefore they are semiclassically set to zero initially in the
simulations. Their coupling to the Higgs modes will drive their
evolution, giving rise to a rapid (exponential) growth of the GW and
inflaton modes.  Their subsequent non-linear evolution will be well
described by the lattice simulations.

In the next subsections we will describe the different evolution 
stages found in our simulations.

\subsection{Tachyonic growth}

In this subsection we will compare the analytical estimates with our
numerical simulations for the initial tachyonic growth of the Higgs
modes and the subsequent growth of gravitational waves. The first
check is that the Higgs modes grow according to paper I.
There we found that 
\begin{equation}\label{kphi2}
\indent 
k|\phi_k(t)|^2 \simeq v^2\,A(\tau)\,e^{-B(\tau)k^2}\,, 
\end{equation}
with $A(\tau)$ and $B(\tau)$ are given in paper I,
\begin{equation}\label{ABT}
\indent 
A(\tau) = {\pi^2(1/3)^{2/3}\over2\Gamma^2(1/3)}\,{\rm Bi}^2(\tau)\,,
\hspace{5mm}
B(\tau) = 2(\sqrt\tau - 1)\,,
\end{equation}
which are valid for $\tau>1$, and where ${\rm Bi}(z)$ is the Airy 
function of the second kind. Indeed, we can
see in Fig.~\ref{HigSpec} that the initial growth, from $mt=6$ to
$mt=10$, follows precisely the analytical expression, once taken into
account that in Eq.~(\ref{kphi2}) the wavenumber $k$ and time $\tau$ are
given in units of $M=(2V)^{1/3}m$.

\begin{figure}[t]
\begin{center}
\includegraphics[width=5.5cm,height=8.5cm,angle=-90]{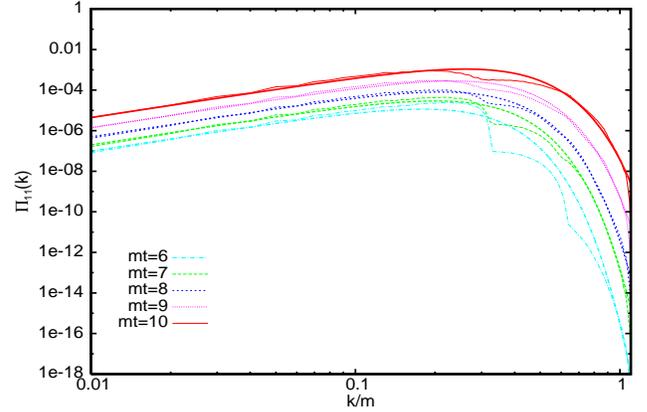}
\end{center}
\vspace*{-5mm}
\caption{The Fourier transform of the anisotropic stress tensor.  We
compare the numerical simulations of $\Pi_{11}(k,t)$ for $p_{\rm min}
= 0.01$ with the analytical expressions (dashed lines) for $mt=5 -
10$, i.e. during the tachyonic growth. The small deviations at $k\leq
m$ are simulation artifacts due to the initial UV cut-off
implementation and soon disappear.}
\label{PijSpec}
\vspace*{-3mm}
\end{figure}

The comparison between the tensor modes $h_{ij}(k,t)$ and the
numerical results is somewhat more complicated. We should first
compute the effective anisotropic tensor T$_{ij}({\bf
k},t)$~(\ref{GWfakeSource}) from the gradients of the Higgs field (those
of the inflaton are not relevant during the tachyonic growth), as
follows,
\begin{equation}
\indent 
\tilde\Pi_{ij}({\bf k},t) = \int {d^3{\bf x}\,e^{-i{\bf k}{\bf x}}
\over(2\pi)^{3/2}}
\left[\nabla_i\phi^a\,\nabla_j\phi^a({\bf x},t) 
\right]\,,
\end{equation}
where
\begin{equation}
\indent 
\nabla_i\phi^a({\bf x},t) = \int {d^3{\bf q}\over(2\pi)^{3/2}}
\,iq_i\,\tilde{\phi^a}({\bf q},t)\,e^{-i{\bf q}{\bf x}}\,.
\end{equation}
After performing the integral in ${\bf x}$ and using the delta
function to eliminate ${\bf q}'$, we make a change of variables
${\bf q}\to {\bf q}+{\bf k}/2$, and integrate over ${\bf q}$,
with which the Fourier transform
of the anisotropic stress tensor becomes 
\begin{eqnarray}\label{Pijk}
\tilde\Pi_{ij}({\bf k},t) = 
k_i\,k_j\,{A(\tau)\over B(\tau)\sqrt2}\,
\Psi\left[{1\over2},0;{B(\tau)k^2\over4}\right]
\,e^{-{1\over4}B(\tau)k^2}, \nonumber \\
\end{eqnarray}
which gives a very good approximation to the numerical results, see
Fig.~\ref{PijSpec}, with $\Psi(1/2,0;z)\simeq(\pi^{-1}+z)^{-1/2}$ 
being the Kummer function.

Finally, with the use of $\tilde\Pi_{ij}({\bf k},t)$, we can compute 
the tensor fields,
\begin{eqnarray}
\hspace{3mm}
h_{ij}({\bf k},t)&=&(16\pi G)\,\int_0^t\,dt'\,{\sin k(t-t')\over k}\,
\tilde\Pi_{ij}\\
\hspace{3mm}
\partial_0 h_{ij}({\bf k},t)&=&(16\pi G)\,\int_0^t\,dt'\,\cos k(t-t')\,
\tilde\Pi_{ij}.
\end{eqnarray}
Using the analytic expression in Eq.~(\ref{Pijk}) one can perform the
integrals and obtain expressions that agree surprisingly well with
the numerical estimates. This allows one to compute the density in
gravitational waves, $\rho_{_{\rm GW}}$, at least during the initial 
tachyonic stage in terms of analytical functions, and we reproduce
the numerical results, see Fig.~\ref{fig4}.

We will now compare our numerical results with the analytical estimates.
The tachyonic growth is dominated by the faster than exponential
growth of the Higgs modes towards the true vacuum.
The (traceless) anisotropic strees tensor $\Pi_{ij}$ grows rapidly to
a value of order $k^2|\phi|^2 \sim 10^{-3}\,m^2v^2$, which gives a
tensor perturbation 
\begin{equation}
\indent
\left|h_{ij} h^{ij}\right|^{1/2} \sim 16\pi
G v^2 (m\Delta t)^2 10^{-3}\,,
\end{equation}
and an energy density in GW, 
\begin{equation}
\indent
\rho_{_{\rm GW}}/\rho_0 \sim 64\pi G v^2\,(m\Delta t)^2 10^{-6} \sim Gv^2\,, 
\end{equation}
for $m\Delta t \sim 16$.  In the case at hand, with $v=10^{-3}\,M_P$, we
find $\rho_{_{\rm GW}}/\rho_0 \sim 10^{-6}$ at symmetry breaking, which
coincides with the numerical simulations at that time, see
Fig.~\ref{fig3}. 

As shown in Ref.~\cite{symmbreak}, the spinodal instabilities grow
following the statistics of a gaussian random field, and therefore one
can use the formalism of~\cite{BBKS} to estimate the number of peaks
or lumps in the Higgs spatial distribution just before symmetry
breaking.  As we will discuss in the next section, these lumps will
give rise via non-linear growth to lump invagination and the formation
of bubble-like structures with large density gradients, expanding at
the speed of light and colliding among themselves giving rise to a
large GWB. The size of the bubbles upon collision is essentially
determined by the distance between peaks at the time of symmetry
breaking, but this can be computed directly from the analysis of
gaussian random fields, as performed in Ref.~\cite{symmbreak}.

This analysis works only for the initial (linear) stage before symmetry
breaking.  Nevertheless, we expect the results to extrapolate to later
times since once a bubble is formed around a peak, it remains there at
a fixed distance from other bubbles. This will give us an idea of the
size of the bubbles at the time of collision.

We estimate the number density of peaks as~\cite{BBKS}
\begin{equation}
\indent n_{\rm peak}(\tau) = {2\over3\sqrt3\,\pi^2}\xi_0(\tau)^{-3}(\nu^2-1)
\exp[-\nu^2/2]\,, 
\end{equation}
where $\nu=\phi_c/\sigma(\tau)$ corresponds to the ratio of the field
threshold $\phi_c$ over the dispersion
\begin{equation}
\indent \sigma(\tau) = {\sqrt\lambda\over\pi}(2V)^{1/3}\Big({A(\tau)\over 
B(\tau)}\Big)^{1/2}\,,
\end{equation}
with $A(\tau)$ and $B(\tau)$ given in Eq.~(\ref{ABT}). The average
size of the gaussian lumps is $\xi_0(\tau)=2B^{1/2}(\tau)\,m^{-1}$,
where time is given in units $\tau = (2V)^{1/3}\,mt$, see
Ref.~\cite{symmbreak}

The distance between peaks can be estimated as twice the radius of the
average bubble, with volume $V_{\rm peak} = 4\pi/3\,R_{\rm peak}^3$.
Since the total volume $L^3$ is divided into $N_{\rm peak}$ bubbles,
we find 
\begin{equation}
\indent d_{\rm peak} = 2R_{\rm peak} = {1\over ma}\,\Big({6\over\pi
n_{\rm peak}}\Big)^{1/3}\,a,
\end{equation}
which is typically of order 30 to 40 lattice units, for $\phi_c \simeq
0.5 - 0.8$, $V=0.024$ and $\lambda=0.125$, with lattices sizes given
by $p_{\rm min}=0.15\,m$ and $N=128$.

What is interesting is that decreasing either $\lambda$ or $V$, the
distance between initial lumps increases and thus also the size of the
final bubbles upon collision. As we will show in the next section, the
amplitude of GW depends on the bubble size squared, and therefore it
is expected that for lower lambda we should have larger GW amplitude.
We have not seen, however, such an increase in amplitude, but a 
detailed analysis is underway and will be presented elsewhere.

\subsection{Bubble collisions}

The production of gravitational waves in the next stage proceeds
through `bubble' collisions. In Ref.~\cite{tachyonic} we showed
explicitly that symmetry breaking is not at all a homogeneous
process. During the breaking of the symmetry, the Higgs field develops
lumps whose peaks grow up to a maximum value $|\phi|_{\rm max}/v =
4/3$ (see paper I), and then decrease creating approximately
spherically symmetric bubbles, with ridges that remain above $|\phi| =
v$. Finally, neighboring bubbles collide and the symmetry gets further
broken through the generation of higher momentum modes.  Since
initially only the Higgs field sources the anisotropic stress-tensor
$\Pi_{ij}$, then we expect the formation of structures (see section
IV.A) in the tensor metric perturbation, correlated with the Higgs
lumps. The dependence of the $h_{ij}$ tensor on the gradient of the
Higgs field, see~Eq.(\ref{GWeq}), is responsible of the formation of
those structures in the energy density spatial distribution of the
GWB.

In section V of this paper we will give account of the explicit form
of the structures developed in the spatial distribution of
$\rho_{_{\rm GW}}$ related with the first collisions among the
bubble-like structures of the Higgs field. We will present
simultaneously the evolution of both the Higgs' spatial distribution
when the first bubbles start colliding, and of the corresponding
structures in the GW energy density $\rho_{_{\rm GW}}$. We leave for a
forthcoming publication the details of an analytical formalism
describing the formation and subsequent evolution of such GW
structures. In this sub-section we will just give an estimate of the
burst in GW produced by the first collisions of the Higgs bubble-like
structures.

Thus, as described in Ref.~\cite{KosowskyTurner} for the case of first
order phase transitions, which involves the collision of vacuum
bubbles, we can give a simple estimate of the order of magnitude of
the energy fraction radiated in the form of gravitational waves when
two Higgs bubble-like structures collide. In general, the problem of
two colliding bubbles has several time and length scales: the duration
of the collision, $\Delta t$; the bubbles' radius $R$ at the moment of
the collision; and the relative speed of the bubble walls. In section
IV.B we found that the typical size of bubbles upon collisions, is of
the order of $R\approx 10m^{-1}$, while the growth of the bubble's
wall is relativistic. Then we can assume than the time scale
associated with bubble collisions is also $\Delta t\,\sim\,R$.
Assuming the bubble walls contain most of the energy density, and
since they travel close to the speed of light, see paper I, it is 
expected that the asymmetric collisions will copiously produce GW.

Far from a source that produces gravitational radiation, the dominat
contribution to the amplitude of GW is given by the acceleration of
the quadrupole moment of the Higgs field distribution. Given
the energy density of the Higgs field, $\rho_{\rm H}$, we can compute
the (reduced) quadrupole moment of the Higgs field spatial
distribution, $Q_{ij} = \int d^3x\,(x_i x_j -
x^2\delta_{ij}/3)\,\rho_{\rm H}(x)$, such that the amplitude of the
gravitational radiation, in the TT gauge, is given by 
$h_{ij}\,\sim\,(2G/r)\ddot Q_{ij}$. A significant amount of energy can be
emitted in the form of gravitational radiation whenever the
quadrupole moment changes significantly fast: through the bubble
collisions in this case. The power carried by these waves can be
obtained via~(\ref{rhoTotal}) as
\begin{eqnarray}
\indent P_{_{\rm GW}} = \frac{G}{8\pi}\int d\Omega\,\left\langle  
\dddot{Q}_{ij}\dddot{Q}^{ij}\right\rangle\,. 
\end{eqnarray} 
Omitting indices for simplicity, as the power emitted in gravitational
waves in the quadrupole approximation is of order $P_{_{\rm
GW}}\,\sim\,G(\dddot{Q})^2$, while the quadrupole moment is of order
$Q \sim R^5\rho_{\rm H}$, we can estimate the power emitted in GW
upon the collision of two Higgs bubbles as
\begin{equation}
\indent P_{_{\rm GW}}\,\sim\,G\left(\frac{R^5\rho}{R^3}\right)^2\,
\sim\,G\rho_{\rm H}^2\,R^4
\end{equation} 
The fraction of energy density carried by these waves, $\rho_{_{\rm
GW}}\,\sim\,P_{_{\rm GW}}\Delta t/R^3\,\sim\,P_{_{\rm
GW}}/R^2\,\sim\,G\rho_{\rm H}^2\,R^2$, compared to that of the initial
energy stored in the two bubble-like structures of the Higgs field,
will be $\rho_{_{\rm GW}}/{\rho_{\rm H}} = G\rho_{\rm H}R^2$.  Since
the expansion of the universe is negligible during the bubble
collision stage, the energy that drives inflaton, $\rho_0\,\sim\,
m^2v^2$, is transferred essentially to the Higgs modes during
preheating, within an order of magnitude, see Fig.~\ref{fig3}. Thus,
recalling that $R\,\sim 10m^{-1}$, the total fraction of energy in GW
produced during the bubble collisions to that stored in the Higgs
lumps formed at symmetry breaking, is given by
\begin{eqnarray}\label{GWbubbles}
\indent \frac{\rho_{_{\rm GW}}}{\rho_0}\,\sim 0.1\,G\rho_0\,R^2\,
\sim\,(v/M_p)^2\,,
\end{eqnarray} 
giving an amplitude which is of the same order as is observed in the
numerical simulations, see Fig.~\ref{fig3}.  Of course, an exhaustive
analytical treatment of the production of GW during this stage of
bubble collisions remains to be done, but we leave it for a future
publication.

\begin{figure}[t]
\begin{center}
\includegraphics[width=5.5cm,height=8.5cm,angle=-90]{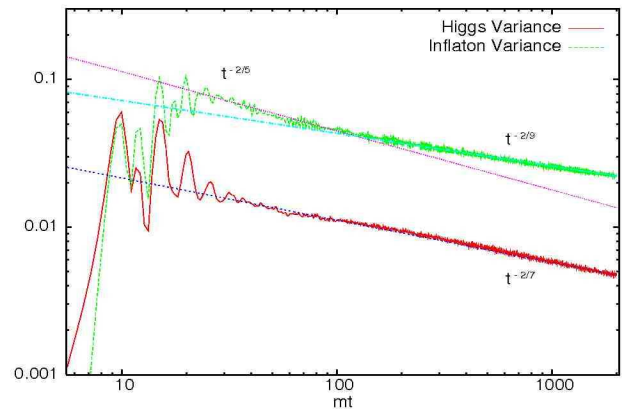}
\end{center}
\vspace*{-5mm}
\caption{Variance of the Inflaton and the Higgs field as a function of
time, the former normalized to its critical value, the latter
normalized to its \textit{v.e.v.}. As expected in a turbulent regime,
these variances follow a power law $\sim t^{-2p}$ with $p$ a certain
critical exponent, although the slope of the Inflaton's variances
evolves in time. The curves are produced from an average over 10 
different statistical realizations.}
\label{variances_fig}
\vspace*{-3mm}
\end{figure}

\subsection{Turbulence}

The development of a turbulent stage is expected from the point of
view of classical fields, as turbulence usually appears whenever there
exists an active (stationary) source of energy localized at some scale
$k_{\rm in}$ in Fourier space. As first pointed out by
Ref.~\cite{MichaTkachev}, in reheating scenarios the coherently
oscillating inflaton zero-mode plays the role of the pumping-energy
source, acting at a well defined scale $k_{\rm in}$ in Fourier space,
given by the frequency of the inflaton oscillations. Thus, the
inflaton zero-mode pumps energy into the rest of the fields that
couple to it as well as into the non-zero modes of the inflaton field
itself. Apart from $k_{\rm in}$, there is no other scale in Fourier
space where energy is accummulated, dissipated and/or infused. So, as
turbulence is characterized by the transport of some conserved
quantity, energy in our case, we should expect a flow of energy from
$k_{\rm in}$ towards higher (direct cascade) or smaller (inverse
cascade) momentum modes. In typical turbulent regimes of classical
fluids, there exits a sink in Fourier space, corresponding to that
scale at which the (direct) cascade stops and energy gets
dissipated. However, in our problem there is no such sink so that the
transported energy cannot be dissipated, but instead it is used to
populate high-momentum modes. For the problem at hand, there exists a
natural initial cut-off $k_{\rm out} \sim \lambda^{1/2}v,$ such that
only long wave modes within $k<k_{\rm out}$, develop the spinodal
instability. Eventually, after the tachyonic growth has ended and the
first Higgs' bubble-like structures have collided, the turbulent
regime is established. Then the energy flows from small to greater
scales in Fourier space, which translates into the increase of $k_{\rm
out}$ in time.

\begin{figure}[t]
\begin{center}
\includegraphics[width=5.5cm,height=8.5cm,angle=-90]{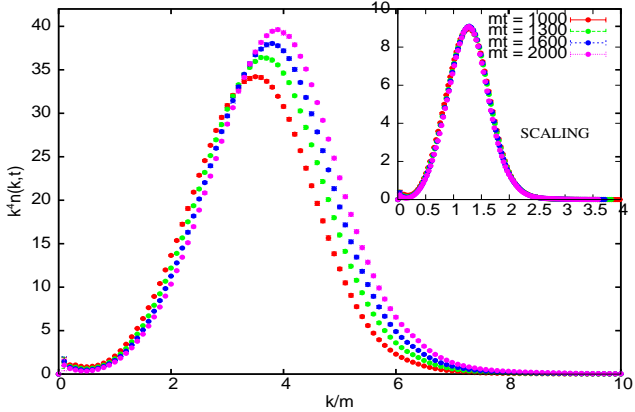}
\end{center}
\vspace*{-5mm}
\caption{Some snapshots of the evolution of the spectral particle
occupation numbers of the Higgs field at different times, each
averaged over 10 statistical realizations. We multiply them by $k^4$
so we can see better the scaling behaviour. In the upper right corner,
we plot the inverse relation of~(\ref{selfSimilar}), $n_0(k t^{-p})
= t^{\gamma p}n(k,t)$, also averaged over 10 realizations for each
time. The scaling behaviour predicted by wave kinetic turbulent
theory~\cite{MichaTkachev}, is clearly verified.}
\label{higgsScaling_fig}
\vspace*{-3mm}
\end{figure}

\begin{figure}[t]
\begin{center}
\includegraphics[width=5.5cm,height=8.5cm,angle=-90]{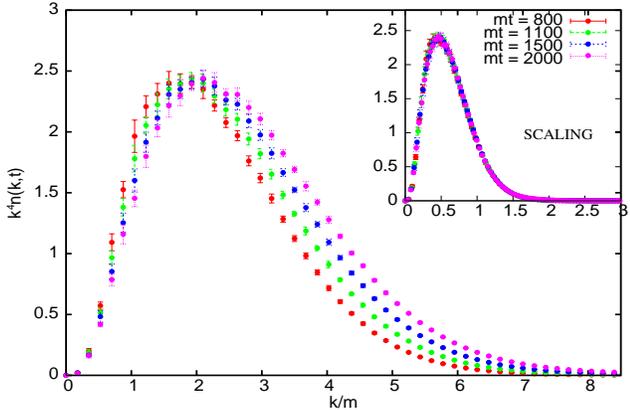}
\end{center}
\vspace*{-5mm}
\caption{Different times of the evolution of the particle occupation
numbers spectra of the Inflaton, multiplied by $k^4$, and averaged
over 10 statistical realizations or each time. Again, in the upper
right corner, we plot the inverse relation of~(\ref{selfSimilar}),
$n_0(k) = t^{\gamma p}n(kt^{p},t)$, also averaged over 10
realizations for each time.}
\label{inflatonScaling_fig}
\vspace*{-3mm}
\end{figure}

In Ref.~\cite{magnetic} we already accounted for the turbulent stage
reached in a hybrid model with gauge fields. However, we don't
consider gauge fields here, so the number of degrees of freedom is
different from that of Ref.~\cite{magnetic} and, therefore, the turbulent
dynamics of the Inflaton and the Higgs fields should be different. In
particular, when the turbulence has been fully established, if the
wave (kinetic) turbulence regime of the fields' dynamics is valid, the
time evolution of the variance of a turbulent field $f({\bf x},t)$,
should follow a power-law-like scaling~\cite{MichaTkachev}
\begin{equation}
\label{variances}
\indent \mathrm{Var}(f(t)) = \left\langle f(t)^2\right\rangle -
\left\langle f(t)\right\rangle^2 \propto \ t^{-2p} \,,
\end{equation}
with $p = 1/(2N-1)$ and $N$ the number of scattering fields in a
`point-like collision'. In fact, such time behaviour corresponds only
to the case of the so called \textit{free turbulence}, when the energy
stored in the pumping source is subdominant to the energy in the
turbulent fields. In our case, this condition is reached very soon
after the symmetry breaking, so we don't expect a significant stage of
\textit{driven turbulence}, which would make the variance to increase
(Only the inflaton seems to increase its variance between $mt=10$ and
$mt=30$, but it is not very pronounced). In Fig.~\ref{variances_fig}
we have plotted the time evolution of the variances of the Inflaton
$\chi$ and of the Higgs modulus $\phi = \sqrt{\sum_a\phi_a^2}$, and
fitted the data with a power-law like~(\ref{variances}), obtaining
\begin{flushleft}
\begin{tabular}{rrll}
\vspace{.2cm} \,\,\,\, & Inflaton: & $p^{-1}_{_I} = 5.1 \pm 0.2$, &
\,\,\,[35:85] \\ \vspace{.2cm} \,\,\,\,& Inflaton: & $p^{-1}_{_I} =
9.03 \pm 0.03$, & \,\,\,[350:2000] \\ \vspace{.1cm} \,\,\,\, & Higgs:
& $p^{-1}_{_H} = 7.02 \pm 0.01$, & \,\,\,[50:2000]
\end{tabular}
\end{flushleft}
where the last brackets on the right correspond to the range in time 
(in units of $m^{-1}$) for which we fitted the data. As can be seen in
Fig.~\ref{variances_fig}, the slope of the Higgs field (in
logarithmic scale), $2p_{_H}\,\sim\,2/7$, remains approximately
constant in time, corresponding to a 4-field dominant interaction.
However, the slope of the Inflaton's variance increases in time, 
i.e. the critical exponent $p_{_I}$ of the Inflaton decreases, 
until it reaches a stationary stage at $mt\sim100$. 
Since $p_{_I}$ is related to the number $N$ of fields interacting in a
collision, if there was a change from one dominant multi-field
interaction to another, this should produce a time-dependent effective
$p_{_I}$, as seen in Fig.~\ref{variances_fig}. However, we will not
try to explain here the origin of such an effective critical exponents
as extracted from the simulations. We will just stress that we 
have checked the robustness of those values under different lattice 
configurations ($N,p_{\rm min}$) and different statistical realizations, 
discarding this way a possible lattice artefact effect. 
As we will see, the critical exponents $p$ determines the speed with 
which the turbulent particle distribution moves over momentum space, 
so this is a crucial parameter. Moreover, although both the classical 
modes of the Inflaton
and the Higgs contribute to the production of GW, the Inflaton's
occupation numbers decrease faster than those of the Higgs so, after a
given time, only the Higgs' modes remain as the main source of GW.

Actually, when turbulence is developed, it is expected that the
distribution function of the classical turbulent fields, the inflaton
and the Higgs here, follow a self-similar
evolution~\cite{MichaTkachev}
\begin{eqnarray}
\label{selfSimilar}
\indent n(k,t) = t^{-\gamma\,p}n_0(k\,t^{-p})
\end{eqnarray}
with $p$ the critical exponent of the fields' variances and $\gamma$ a
certain factor $\sim O(1)$, which depends on the type of turbulence
developed. It is precisely this way that the exponent $p$ determines
the speed of the particles' distribution in momentum space: given a
specific scale $k_c$ such that, for example, the occupation number has
a maximum, that scale evolves in time as $k_c(t) = k_c(t_0)(t/t_0)^{p}$. 
We have seen that the evolution of the Higgs occupation number follows
Eq.~(\ref{selfSimilar}) with $p \approx 1/7$, as expected from the
Higgs variance, and $\gamma \approx 2.7$. Whereas the evolution of the
Inflaton occupation number follows~(\ref{selfSimilar}) even more
accurately than the Higgs, with an ``effective" exponent $p \approx
1/5$, and $\gamma \approx 3.9$. Since the slope of the inflaton's
variance changes in time, the value of the exponents of the inflaton's 
scaling relation will require further investigation. However, despite 
this time evolution of the Inflaton variance, Eq.~(\ref{selfSimilar})
is very well fulfilled by the Inflaton with the given effective 
exponents. So we can perfectly obtain the universal $n_0(k)$ function 
for the Inflaton as well as for the Higgs.

In Figs.~\ref{higgsScaling_fig} and~\ref{inflatonScaling_fig} we have
plotted the occupation numbers of the Higgs and the Inflaton, also
inverting the relation of Eq.~(\ref{selfSimilar}) in order to extract
the \textit{universal} time-independent $n_0(k)$ functions of each
field. As shown in those figures, the distributions follow the
expected scaling behaviour. However, for the range of interest of $k$,
there are small discrepancies of order 0.1-4\% (depending on $k$)
among the different universal functions $n_0^{(i)}(k)$, as obtained 
inverting Eq.~(\ref{selfSimilar}) at different times $mt_i$. The universal
functions $n_0(k)$ plotted in Figs.~\ref{higgsScaling_fig} 
and~\ref{inflatonScaling_fig} have been obtained from averaging over
ten statistical realizations for each time.

\begin{figure}[t]
\begin{center}
\includegraphics[width=5.5cm,height=8.5cm,angle=-90]{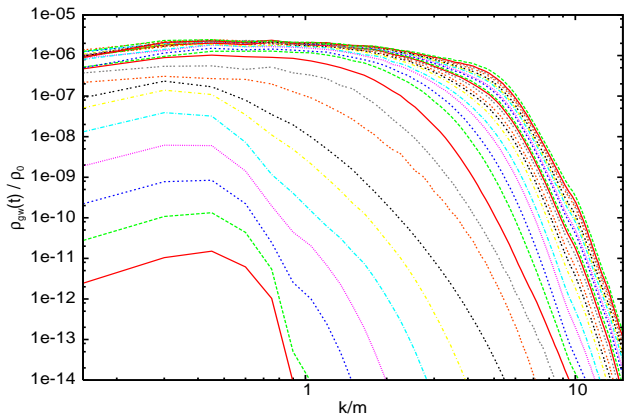}
\end{center}
\vspace*{-5mm}
\caption{Time evolution of the GW spectra from mt = 6 to mt = 2000. The amplitude of the spectra seems to saturate after mt $\sim 100$, although the high momentum tail still moves slowly to higher values of $k$ during the turbulent stage.}
\label{fig10}
\vspace*{-3mm}
\end{figure}

The advantage of the development of a turbulence behaviour is obvious:
it allows us to extrapolate the time evolution of the fields'
distributions till later times beyond the one we can reach with the
simulations. Moreover, the fact that the turbulence develops so early
after the tachyonic instability, also allow us to check for a long
time of the simulation, the goodness of the description of the
dynamics of the fields, given by the turbulent kinetic theory
developed in Ref.~\cite{MichaTkachev}. We have fitted the averaged
universal functions $n_0(k)$ with expressions of the form $k^4\,n_0(k)
= P(k)e^{-Q(k)}$, with $P(k)$ and $Q(k)$ polynomials in $k$, giving: 
\begin{eqnarray}
\begin{array}{lll}
\label{Inflaton_universal}
{\rm Inflaton:} & P(k) = 486.2k^3 & Q(k) = 6.39k \\
\,&\,& \\
\label{Higgs_universal}
{\rm Higgs:} & P(k) = 2.96k^3 & Q(k) = 2.26k^2 - 3.18k
\end{array}
\end{eqnarray} 
There is no fundamental meaning for these expressions, but it is very
useful to have such an analytical control over $n_0(k)$, since this
allows us to track the time-evolution of $n(k,t)$ through
Eq.~(\ref{selfSimilar}). Actually, the classical regime of the
evolution of some bosonic fields ends when the system can be relaxed
to the Bose-Einstein distribution. We are now going to estimate the
moment in which the initial energy density gets fully transferred to
the Higgs classical modes.  Using Eq.(\ref{selfSimilar}) and the
fit~(\ref{Higgs_universal}), we find that the initial energy density 
is totally transfered to the Higgs when (in units $m=1$)
\begin{equation}
\indent
\rho_0 = {1\over4\lambda} =
\int {dk\over k}\,{k^3\over2\pi^2}\,k\,n(k,t) =
{7.565\over2\pi^2}\, t^{(4-\gamma)p}\,,
\end{equation}
where we have assumed that the Higgs' modes have energy $E_k(k,t) =
k\,n(k,t)$. In our case, with $\lambda=1/8$, the conversion of the
initial energy density into Higgs particles and therefore into
radiation is complete by $mt \sim 6\times10^4$. Therefore, if we 
consider this value as a lower bound for the time that classical 
turbulence requires to end, we see that turbulence last for a very 
long time compared to the time-scale of the initial tachyonic and bubbly 
stages. Thus, if GW were significatively sourced during turbulence, 
one should take into account corrections from the expansion of the universe.

In Fig.~\ref{fig10}, we show the evolution of the GW spectra up to
times mt $= 2000$, for a lattice of (N,$p_{\rm min}$) = (128,.15). It
is clear from that figure that the amplitude of the GW saturates to a
value of order $\rho_{gw}/\rho_0 \approx 2\cdot10^{-6}$. At mt
$\approx 50$, the maximum amplitude of the spectra has already reached
$\rho_{gw}/\rho_0 \approx 10^{-6}$, while at time mt $\approx 100$,
the maximum has only grown a factor of 2 with respect to mt $\approx
50$. From times mt $\approx 150$ till the maximum time we reached in
the simulations, mt = 2000, the maximum of the amplitude of the
spectrum does not seem to change significantly, slowly increasing from
$\approx 2\cdot10^{-6}$ to $\approx 2.5\cdot10^{-6}$. Despite this
saturation, we see in the simulations that the the long momentum tail
of the spectrum keeps moving towards greater values. This displacement
is precisely what one would expect from turbulence, although it is
clear that the amplitude of the new high momentum modes never exceed
that of lower momentum. In order to disscard that this displacement
towards the UV is not a numerical artefact, one should further
investigate the role played by the turbulent scalar fields as a source
of GW. Here, we just want to remark that the turbulent motions of the
scalar fields, seem not to increase significatively anymore the total
amplitude of the GW spectrum. Indeed, in a recent paper
\cite{DufauxGW} where GW production at reheating is also considered,
it is stated that GW production from turbulent motion of classical
scalar fields, should be very supressed. That is apparently what we
observe in our simulations although, as pointed above, this issue
should be investigated in a more detailed way. Anyway, here we can
conclude that the expansion of the Universe during reheating in these
hybrid models, does not play an important role during the time of GW
production, and therefore we can be safely ignore it.


\begin{widetext}

\begin{figure}[t]
\centering
\includegraphics[width=7.5cm,height=14.5cm,angle=270]{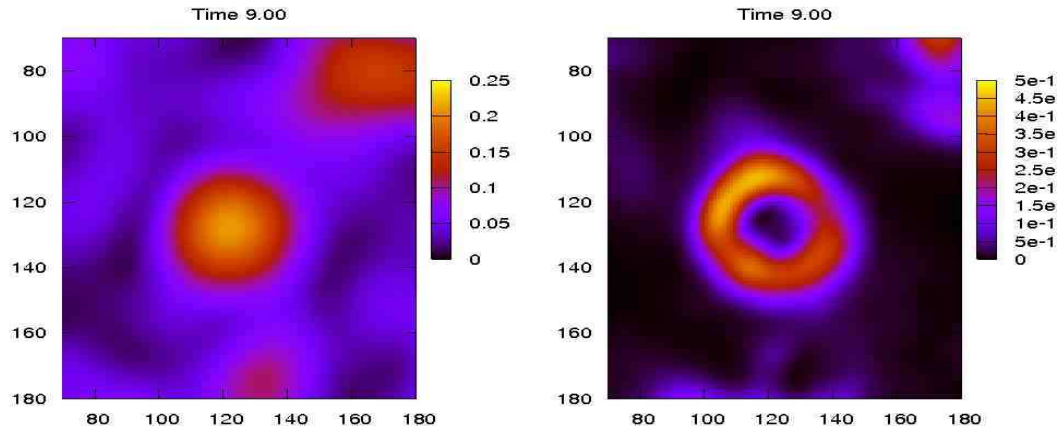}
\caption{Model: $\lambda = 10^{-3}$, $g^2=1$. In the left picture we
show a spatial section of $\left<\phi\right>$. We can see how a
spherical lump is growhing. In the right picture we can see the
structure of the $\rho_{GW}$ in the same place.  A ring is forming
around the Higgs lump. More complex structures are formed in the
regions in which the Higgs bubbles are next, and the GW grow in the
boundary of this lumps, where the gradient of the Higgs and therefore
the $\Pi_{ij}$ tensor grow in this region.}
\label{fig:ring1}
\end{figure}

\end{widetext}

\section{Spatial sections and local GW production}

\begin{figure}[htb]
\vspace{5mm}
\centerline{
\subfigure{
\includegraphics[width=4.5cm,height=9.0cm,angle=-90]{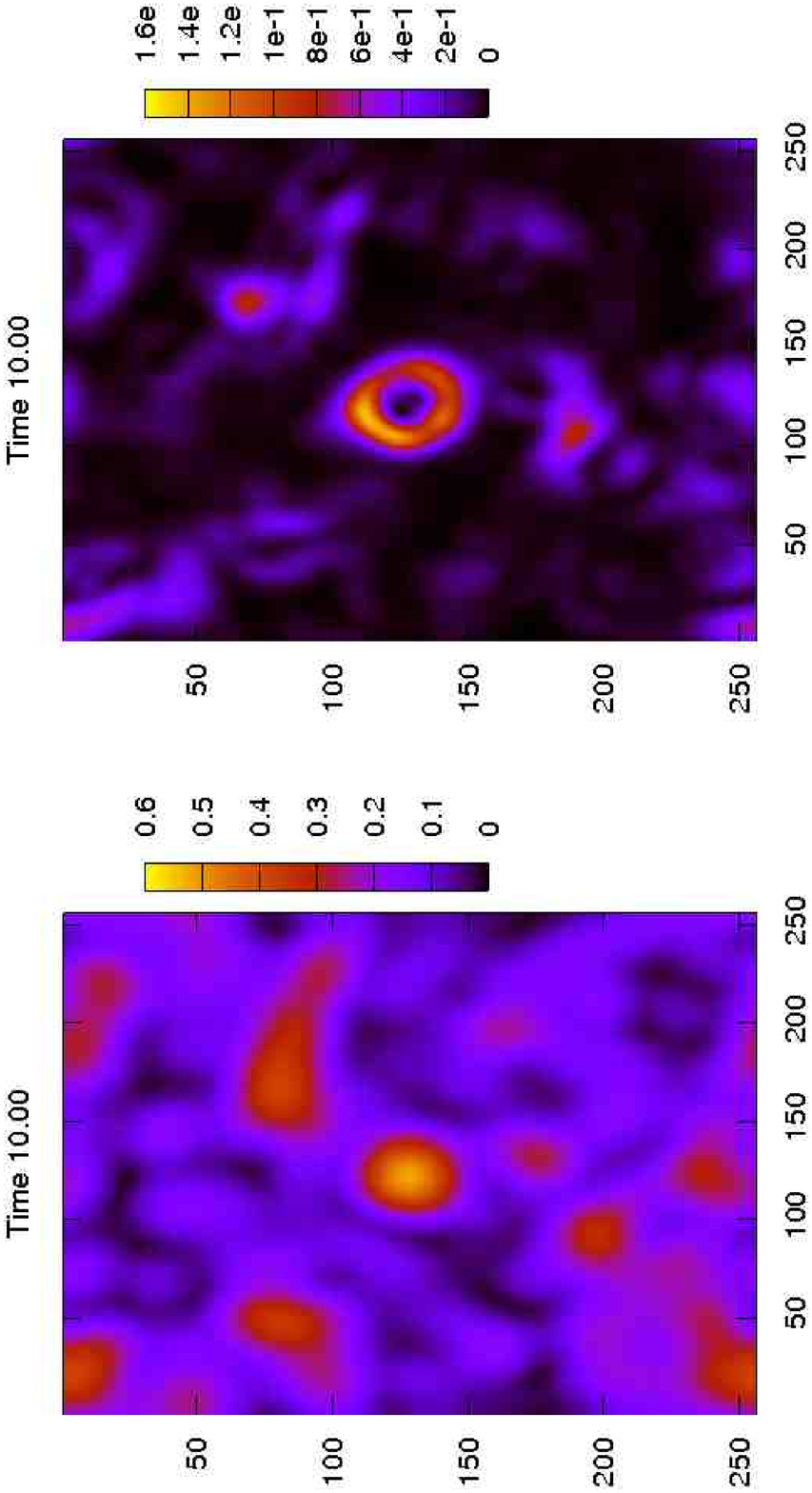}}}
\centerline{
\subfigure{
\includegraphics[width=4.5cm,height=9.0cm,angle=-90]{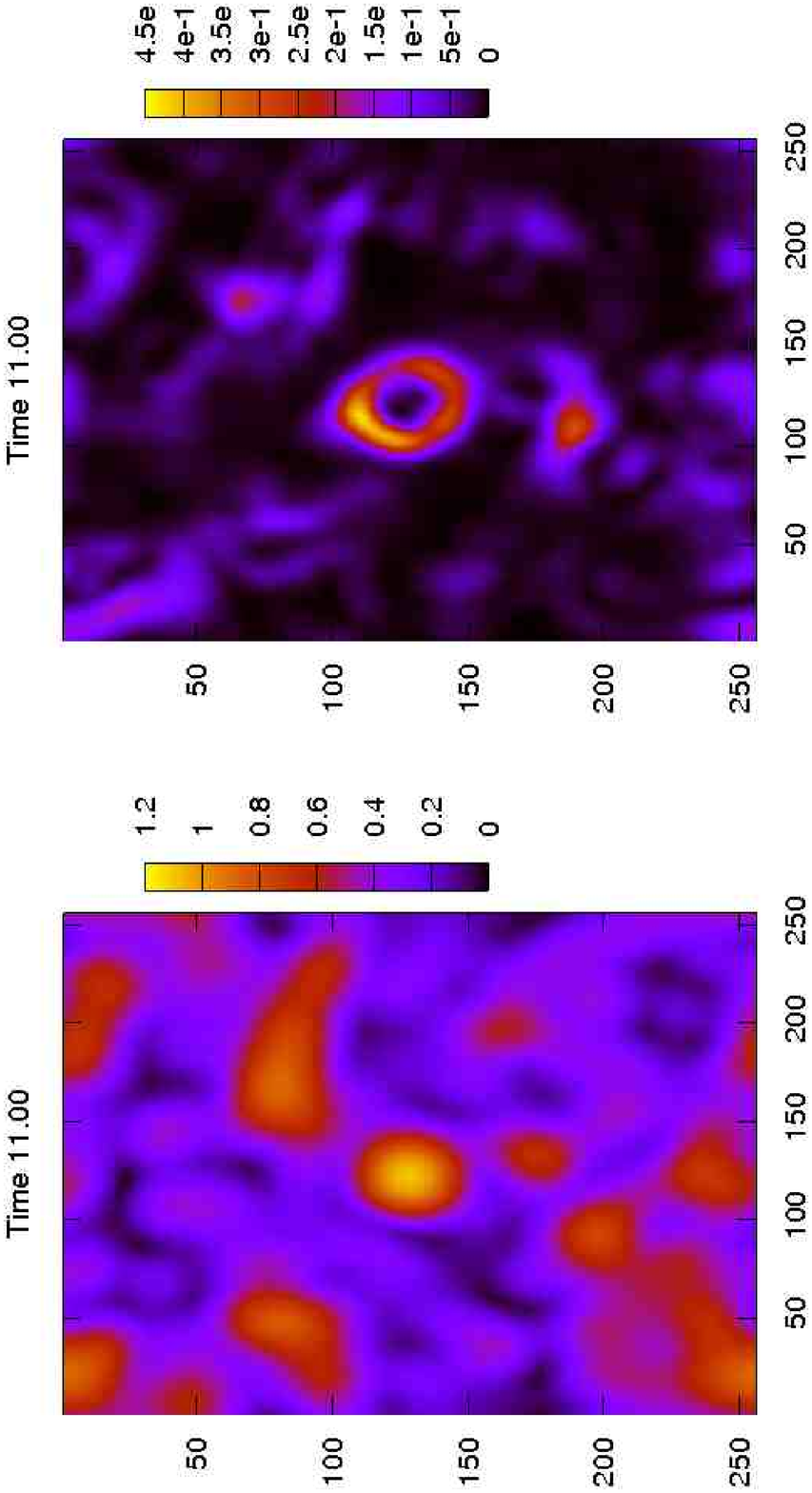}}}
\caption{Here we have got the time evolution of the previous ring
(picture \ref{fig:ring1}) near the symmetry breaking. The bubble is
growing (mt = 10-11), until the symmetry breaking time (mt = 12).}
\label{fig:bubble_expansion1}
\end{figure}

\begin{figure}[htb]
\vspace{5mm}
\centerline{
\subfigure{
\includegraphics[width=4.5cm,height=9.0cm,angle=-90]{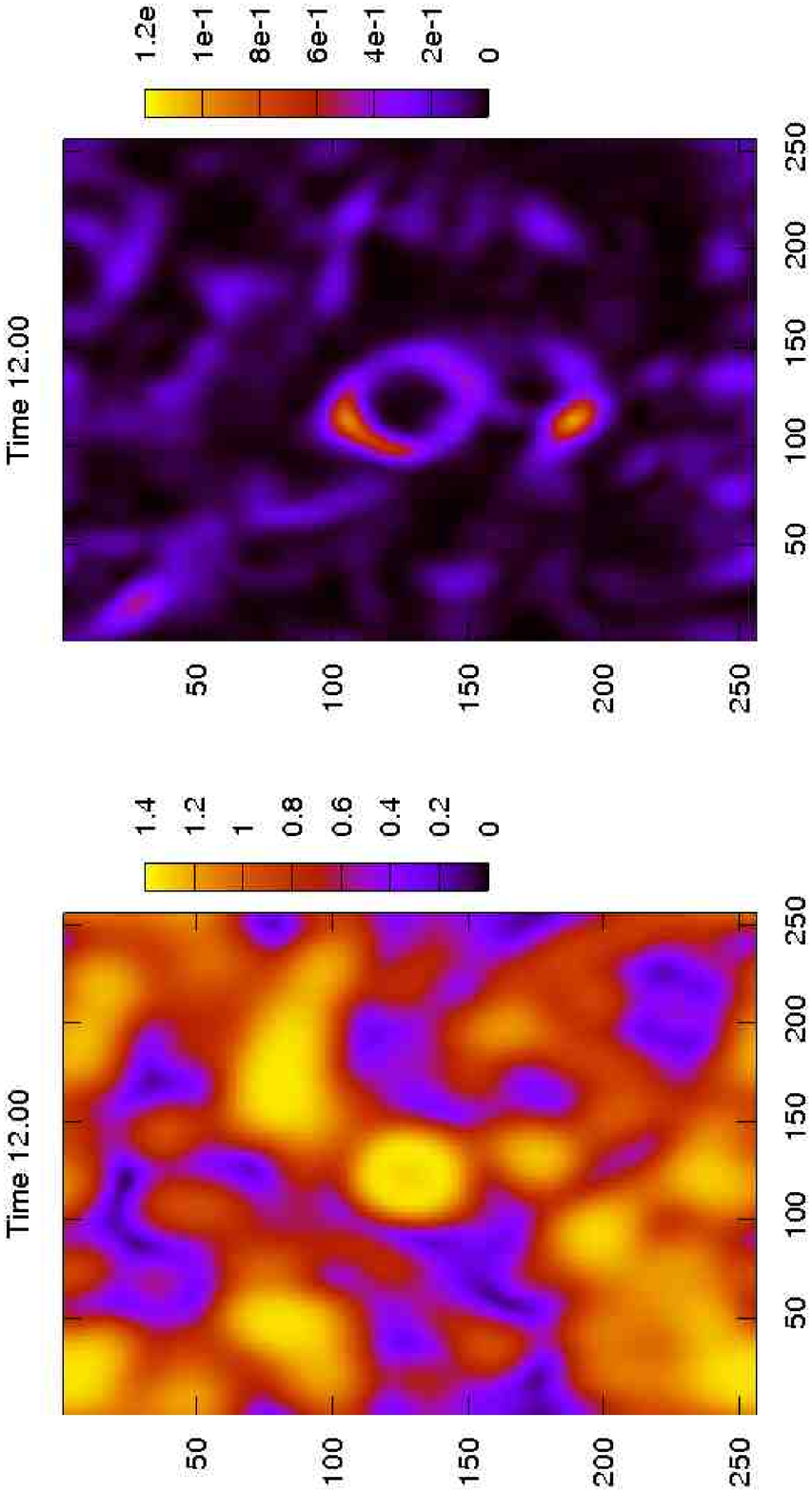}}}
\centerline{
\subfigure{
\includegraphics[width=4.5cm,height=9.0cm,angle=-90]{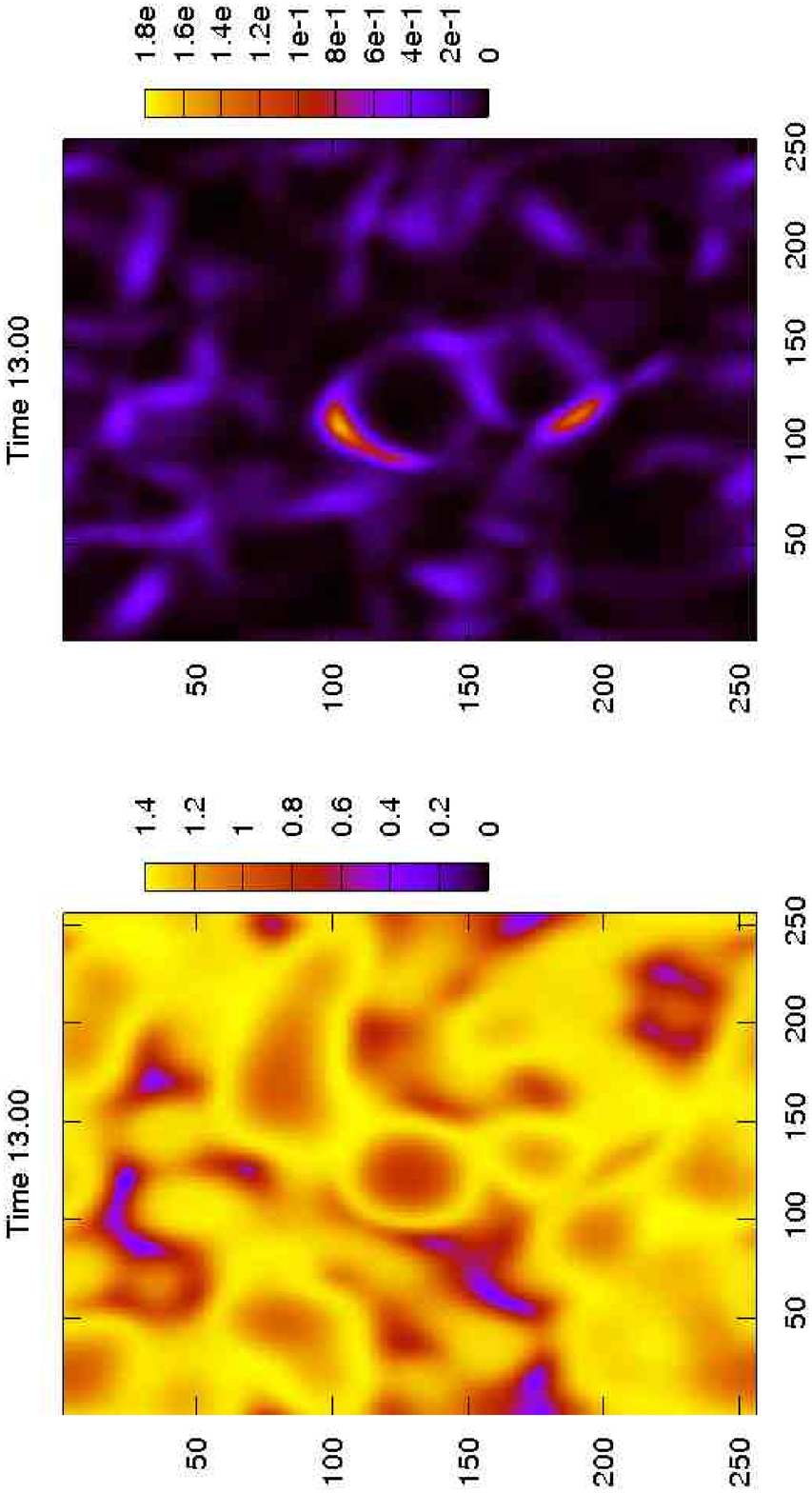}}}
\caption{The Higgs lump begins to invaginate, and the GW ring
expands (mt = 12-13). A similar behavior is observed in a smaller
lump below the biggest Higgs lump, in the same pictures.}
\label{fig:bubble_expansion2}
\end{figure}

In this section, we show a sequence of snapshots ($mt = 5-20$) of the
evolution of the spatial distribution, before the fields are driven to
the turbulent stage. We find that the first stages of the GW dynamics
is strongly correlated with the dynamics of the Higgs oscillations
that give rise to symmetry breaking. A qualitative way of
understanding this question is to analyse the spatial structure of the
$\Pi_{ij}$ tensor, built from spatial gradients of the Higgs and
inflaton fields.  Since the oscillations of $\langle\phi\rangle$ are
due to rapid changes of the Higgs' values in its way of symmetry
breaking, this induces great variations in the behaviour of the
spatial gradients. We are now going to analyse briefly the different
stages showing the most representatives images. Shortly, it will be
available in our web page, a bigger selection of pictures and movies
\cite{LATTICEWEB}.

\begin{figure}[t]
\vspace{5mm}
\centerline{
\subfigure{\includegraphics[width=4.5cm,height=9cm,angle=-90]{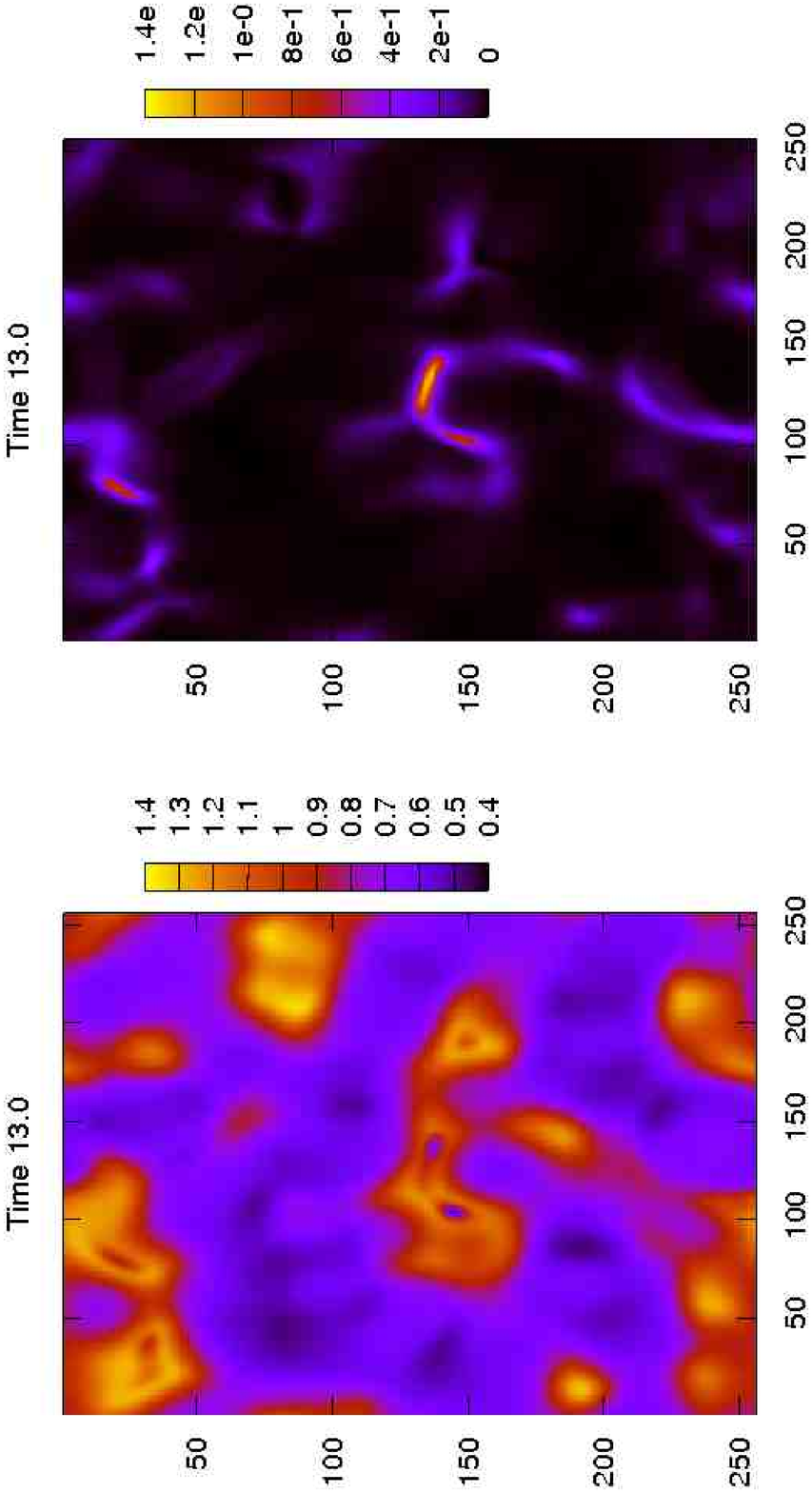}}}
\centerline{
\subfigure{\includegraphics[width=4.5cm,height=9cm,angle=-90]{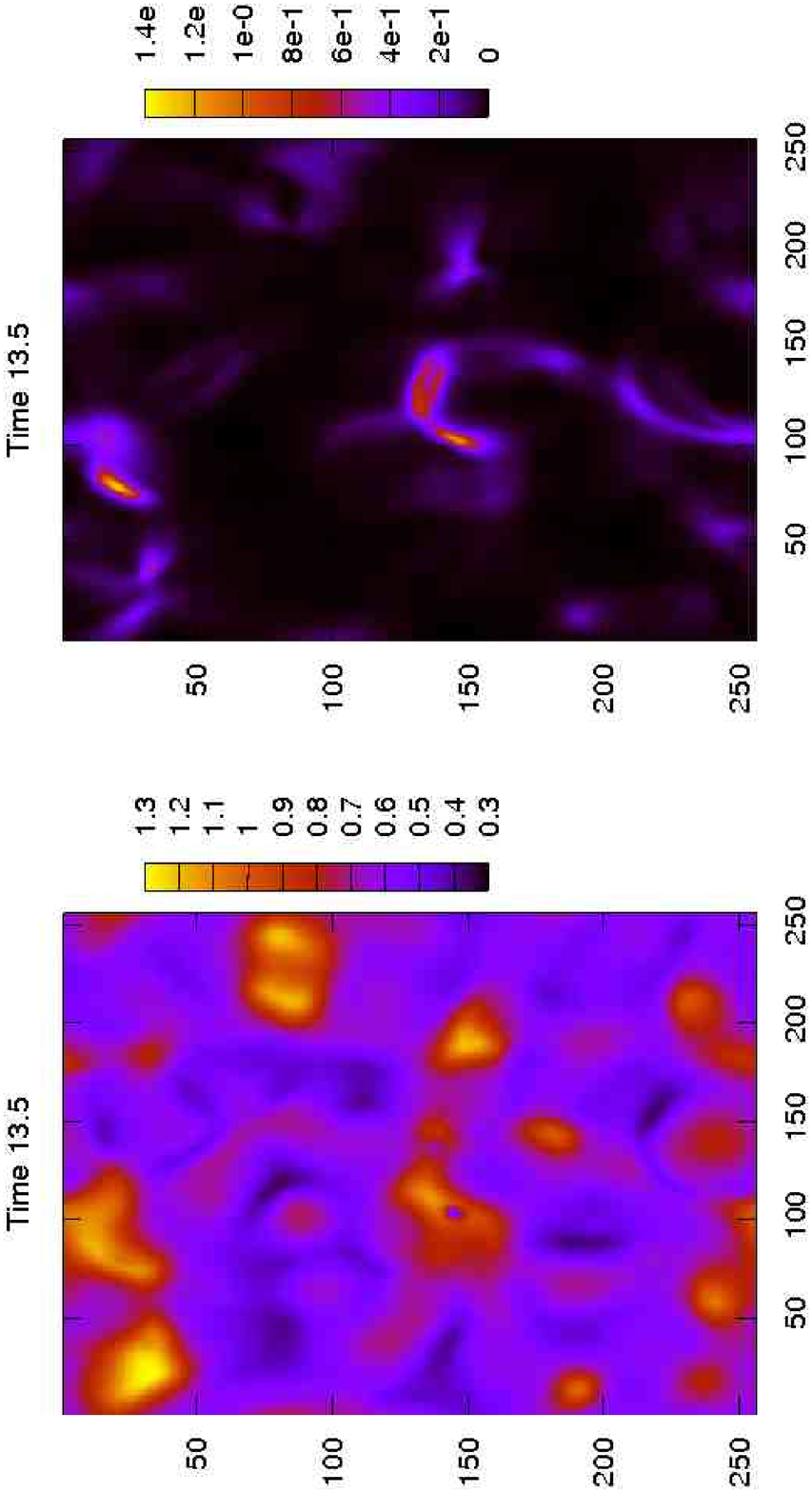}}}
\caption{After symmetry breaking the expansion of Higgs lump
compresses the GW, until the Higgs gradient changes in the first
oscillation (mt =13-13.5).}
\label{fig:compression1}
\end{figure}

\begin{figure}
\vspace{5mm}
\centerline{
\subfigure{\includegraphics[width=4.5cm,height=9cm,angle=-90]{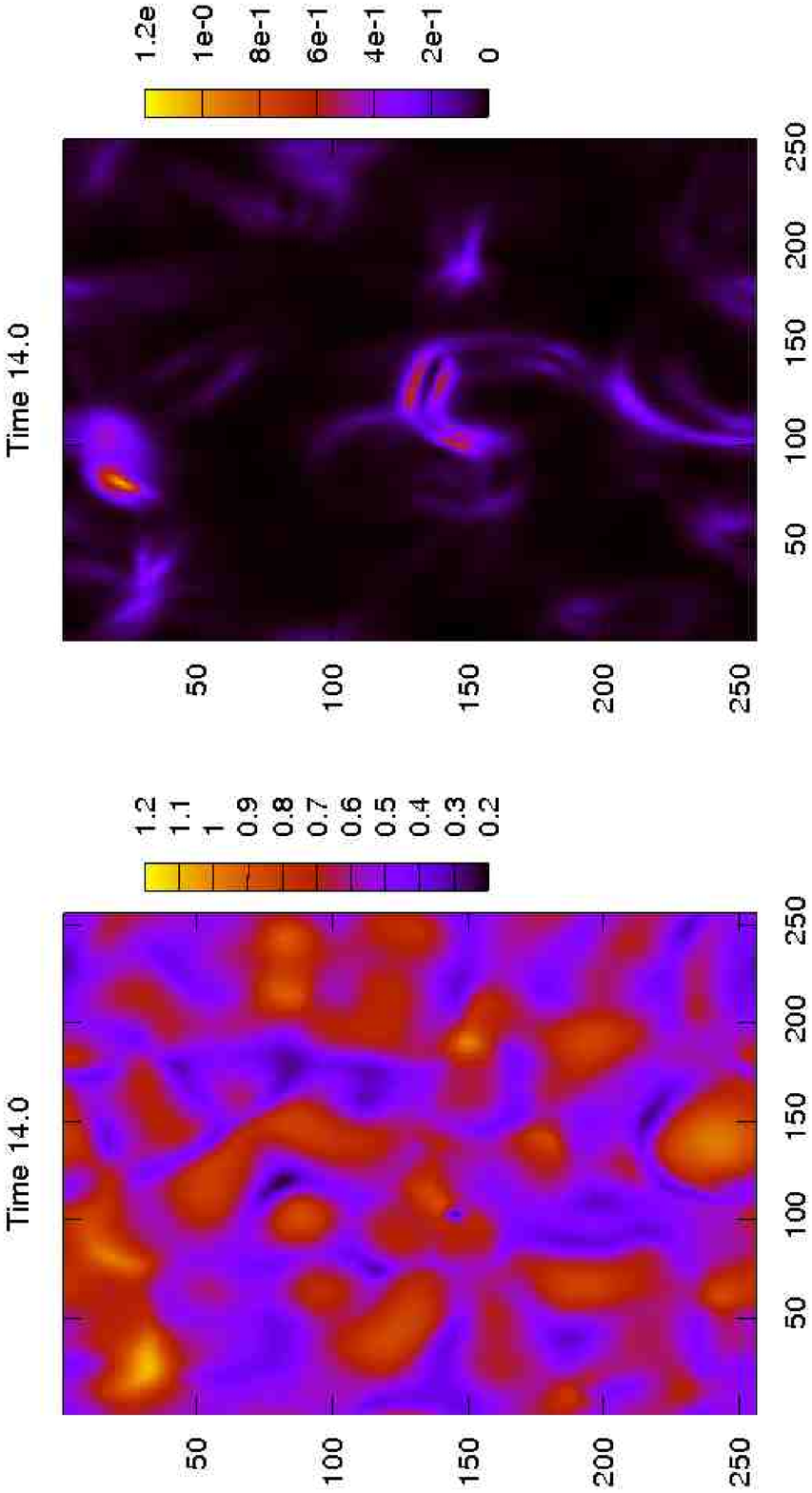}}}
\centerline{
\subfigure{\includegraphics[width=4.5cm,height=9cm,angle=-90]{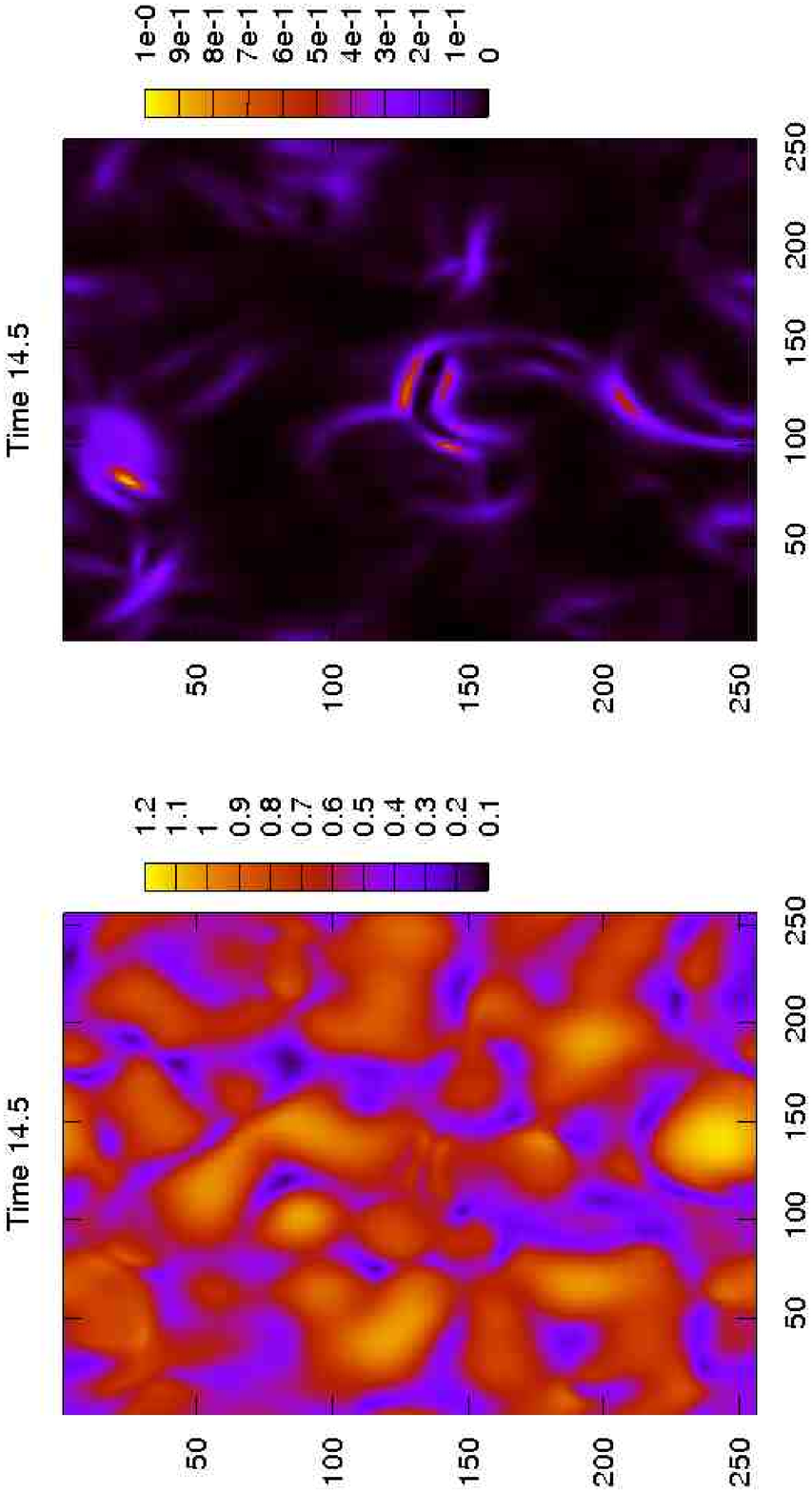}}}
\caption{At this moment, when the Higgs falls, the
GW strutrure is divided in two waves (mt = 14). These wave fronts
propagates in opposite directions (mt = 14.5).}
\label{fig:compression2}
\end{figure}

An interesting conclusion from the set of Figs.~\ref{fig:ring1} $-$
\ref{fig:compression2} is that the Higgs evolution from lump growth,
through invagination to bubble collisions, has a direct translation
into the corresponding growth of gravitational wave energy density.
Not only does the volume-averaged amplitude $\rho_{\rm GW}$ follow
the Higgs time evolution, but the individual local features in the
GWB seem to correspond very closely with the Higgs features.

In the first stage both Higgs and GW backgrounds grow very fast. The
lumps which grow in the Higgs background induce structures around
these, through the gradients appearing in the $\Pi_{ij}$ tensor.  The
geometry of the gravitational structures comes from the position of
the Higgs lump. A typical structure for an isolated lump is a ring of
gravitational waves, see Fig.~\ref{fig:ring1}. More complex structures
can be formed around the position of the Higgs lumps. Before symmetry
breaking these lumps grow according to the previous analysis,
generating domains with a great density of gravitational energy.

The second stage begins when $\langle\phi\rangle = v$ and the symmetry
breaking starts, then the Higgs lumps invaginate and expand, producing
the growth of gravitational waves around of these structures, see
Figs. \ref{fig:bubble_expansion1} and \ref{fig:bubble_expansion2}, one
can see that whenever the bubble walls expand, the variation in the
gradient of the Higgs' field induces the expansion of the GW ring. In
the case of the rings, if the lump is very isolated, the expansion
induces the ring to dilute and disappear, by Gauss law.  In practice,
however, the lumps are never isolated and bubbles collide before the
gradients (and thus the GW) die away.

In the case when two Higgs' bubble-like structures are close by, the
expansion of their walls compresses the GW structures.  This expansion
continues until the first Higgs oscillation, see Fig.~\ref{fig4}. If
the distance between Higgs' structures is small, then the GW can be
diluted, whereas in the other case, a remnant string-like GW structure
survives, and when the Higgs background goes to zero this GW structure
becomes divided into two waves that propagate in opposite directions,
as one can see in Figs.~\ref{fig:compression1} and
\ref{fig:compression2}, which show four snapshots of this process.  A
similar behaviour is observed in the second oscillations.

Finally, the wave fronts propagate, colliding among themselves,
driving the system to the stage of turbulence. We will leave for a
future publication the detailed analysis of the GW production at the
bubble collisions and the subsequent turbulent period.


\section{Gravitational Waves from Chaotic Inflation}

The production of a relic GWB at reheating was first 
addressed by Khlebnikov and Tkachev (KT) in 
Ref.~\cite{TkachevGW}, both for the quadratic and quartic chaotic
inflation scenarios. In these models, the long-wavelength part of the
spectrum is dominated by the gravitational bremsstrahlung associated
with the scattering of the extra scalar particles off the inflaton
condensate, `evaporating' this way the inflaton particles. Using
this fact, KT estimated analytically the amplitude of the power
spectra of GW for the low frequency end of the spectrum,
corresponding to wavelengths of order the size of the horizon at
rescattering. Moreover, KT also studied the GW power spectra
numerically, although just for the massless inflaton case. Recently,
chaotic scenarios were revisited in Ref.~\cite{EastherLim,EastherLim2}, 
accompanied by more precise numerical simulations at different energy 
scales, including the case of a massive inflaton. Finally, also very 
recently, Ref.~\cite{DufauxGW} studied in a very detailed way, 
both analitical and numerically, the evolution of GW produced 
at preheating in the case of a massless inflaton with an extra scalar field.

In Refs.~\cite{TkachevGW} and~\cite{EastherLim}, the procedure
to compute the GW from reheating relied on Weinberg's 
formula for the energy carried by a weak gravitational radiative field
in flat space-time~\cite{Weinberg}. However, in chaotic models, the
expansion of the universe can not be neglected during reheating, 
so Weinberg's formula can only be used in an approximated way, 
if the evolution of the universe
is considered as an adiabatic sequence of stationary
universes. Rescaling fields by a conformal transformation, their
evolution equations can be solved with a numerical integrator, while
the evolution of the scale factor can be calculated
analytically. Discretizing the time, the physical variables can be
recovered from the conformal ones in each time step, thus allowing to
compute the energy of gravitational waves in terms of the physical
fields. In this paper, however, we adopt another approach\footnote{Note that 
Refs.~\cite{EastherLim2} and~\cite{DufauxGW} also work 
in the same theoretical framework, considering the TT tensor 
perturbations on top of a flat FRW space. However, we use a 
different way to extract numerically the GW power spectra, relying on the 
\textit{conmutating procedure}, as detailed
explained in subsection III.B} that 
takes into account expansion of the universe in a
self-consistent manner, and let us calculate at any time 
the energy density and power spectra of the GW produced 
at reheating. As explained in section III and applied to the case of 
hybrid inflation in sections IV and V,
we just solve numerically Eq.~(\ref{GWfakeEq}), together with those 
eqs. of the other Bose
fields and the scale factor, Eqs.~(\ref{inflatonEq}),(\ref{scalarEq}) and 
Eqs.(\ref{hubbleDotEq}),(\ref{hubbleEq}).
Then, using the projector~(\ref{projector}) into the (Fourier transformed) 
solution of Eq.(\ref{GWfakeEq}), we recover the TT d.o.f corresponding 
to GW. This way, we can monitor the total energy
density in GW using Eq.~(\ref{rhoTotal}), or track the evolution 
of the power
spectrum. Using this technique, we will show in this
section that we reproduce, for specific chaotic models, 
similar results to those of other authors.

We adapted the publicly
available LATTICEEASY code~\cite{LatticeEasy}, taking advantage of the
structure of the code itself, incorparating the evolution of 
Eq.~(\ref{GWeq}), together with the equations of the scalar fields, 
Eqs.~(\ref{inflatonEq}) and (\ref{scalarEq}),
into the staggered leapfrog integrator routine. This way, we can solve
at the same time the dynamics of the scalar and tensor fields, within
the framework of an expanding FRW universe~Eqs.(\ref{hubbleDotEq}) and
(\ref{hubbleEq}). 

In particular, we will concentrate only in an scenario with a massless 
inflaton $\chi$, either accompanied or not by an extra scalar field $\phi$.
In the following, we 
will describe the numerical results for GW production at reheating in such 
scenarios, described by the potential
\begin{equation}
\indent V(\chi,\phi) = {\lambda\over 4}\chi^4 + {1\over 2}g^2\chi^2\phi^2
\end{equation}
Rescaling the time by
\begin{eqnarray}
\indent d\tau = {a(\tau)\over a(0)}\chi(0)\sqrt{\lambda}\,dt\,,
\end{eqnarray}
and the physical fields by a conformal transformation as
\begin{eqnarray}
&& \chi_c(\tau) = {a(\tau)\over a(0)}{\chi(\tau)\over\chi(0)}\,, \\
&&\phi_c(\tau) = {a(\tau)\over a(0)}{\phi(\tau)\over\chi(0)}
\end{eqnarray}
then the equations of motion of the inflaton and of the extra scalar field,
Eqs.~(\ref{inflatonEq}) and (\ref{scalarEq}), can be rewritten in
terms of the conformal variables as
\begin{eqnarray}
\label{inflatonEqChao4}
&&\chi_c'' - \nabla^2\chi_c -\frac{a''}{a}\chi_c + 
(\chi_c^2 + q\phi_c^2)\chi_c  = 0 \\
\label{scalarEqChao4}
&&\phi_c'' - \nabla^2\phi_c -\frac{a''}{a}\chi_c + 
q\chi_c^2\phi_c = 0\,,
\end{eqnarray}
where the prime denotes derivative with respect to conformal time. Since
the universe expands as radiation-like in these scenarios,
$a(\tau) \sim \tau$, so the terms proportional to $a''/a$
in Eqs.~(\ref{inflatonEqChao4}) and (\ref{scalarEqChao4}) are soon zero, as
explicitly checked in the simulations. Thanks to this, the
model is conformal to Minkowski.

The parameter $q \equiv g^2/\lambda$ 
controls the strength and width of the resonance.
For the case of a massless inflaton without an extra scalar field,
we just set $q = 0$ in Eq.~(\ref{inflatonEqChao4}) and
ignore Eq.~(\ref{scalarEqChao4}). However, in that case, fluctuations
of the inflaton also grow via parametric resonance. Actually, they
grow as if they were fluctuations of a scalar field coupled to the
zero-mode of the inflaton with effective couplig $q = g^2/\lambda =
3$, see Ref.~\cite{Greene}.

\begin{figure}[htb]
\begin{center}
\includegraphics[width=5.5cm,height=8.5cm,angle=-90]{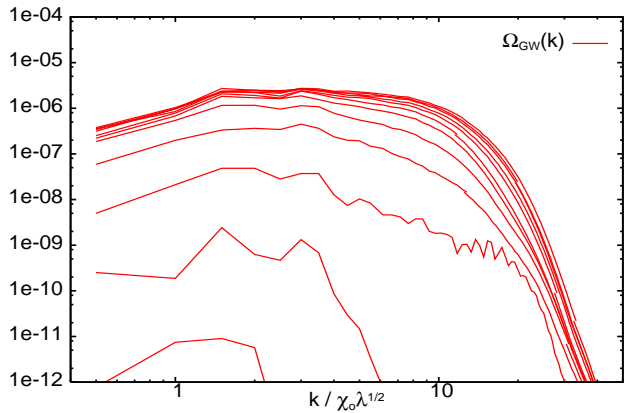}
\end{center}
\vspace*{-5mm}
\caption{The spectrum of the gravitational waves' energy density, 
for coupled case with $\lambda = 10^{-14}$ and $g^2/\lambda = 120$. 
The spectrum is shown accumulated up to different times during GW 
production, so one can see its evolution. At each time, it is normalized 
to the total instant density. This plot corresponds to a N = 128
lattice simulation, from $\tau = 0$ to $\tau = 240$.}
\label{fig6}
\vspace*{-3mm}
\end{figure}

Following Refs.~\cite{TkachevGW} and~\cite{EastherLim}, we set
$\lambda = 10^{-14}$ and $q$ = 120. Since, this case is also 
computed in~\cite{DufauxGW}, we can also compare our results 
with theirs. Moreover, we also present results for the pure 
$\lambda\chi^4$ model with no extra scalar field, a case 
only shown in Ref.~\cite{TkachevGW}.

\begin{figure}[htb]
\begin{center}
\includegraphics[width=5.5cm,height=8.5cm,angle=-90]{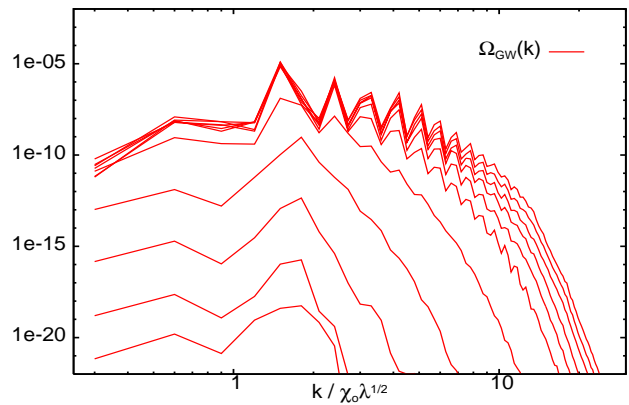}
\end{center}
\vspace*{-5mm}
\caption{The spectrum of the gravitational waves' energy density, 
for the pure case, with $\lambda = 10^{-14}$. Again, we show 
the spectrum accumulated up to different times during GW 
production, normalized to the total instant density at each time.
The plot corresponds to a N = 128 lattice simulation, 
from $\tau = 0$ to $\tau = 2000$.}
\label{fig7}
\vspace*{-3mm}
\end{figure}

We begin our simulations at the
end of inflation, when the homogeneous inflaton verifies $\chi_0
\approx 0.342 M_p$ and $\dot\chi_0 \approx 0$. We took initial quantum
(conformal) fluctuations $1/\sqrt{2k}$ for all the modes up to a
certain cut-off, and only added an initial zero-mode for the inflaton,
$\chi_c(0) = 1$, $\chi_c(0)' = 0$. In Figs.~\ref{fig6} and~\ref{fig7}, 
we show the 
evolution of $\Omega_{_{GW}}$ during reheating, normalized to the 
instant density at each time step, for the coupled and the pure 
case, respectively. In the case with an extra scalar field, 
the amplitude of the GWB saturates at the end of parametric 
resonance, when the fields 
variances have been stabilized. This is the beginnig of the turbulent 
stage in the scalar fields, which seems not to source anymore the 
production of GWs, as already stated in Refs.\cite{EastherLim,DufauxGW}.
For the pure case, we also see the saturation of the amplitude of the 
spectra, see Fig.~\ref{fig7}, although the long momenta tail seems to 
slightly move toward higher values.

\begin{figure}[htb]
\begin{center}
\includegraphics[width=5.5cm,height=8.5cm,angle=-90]{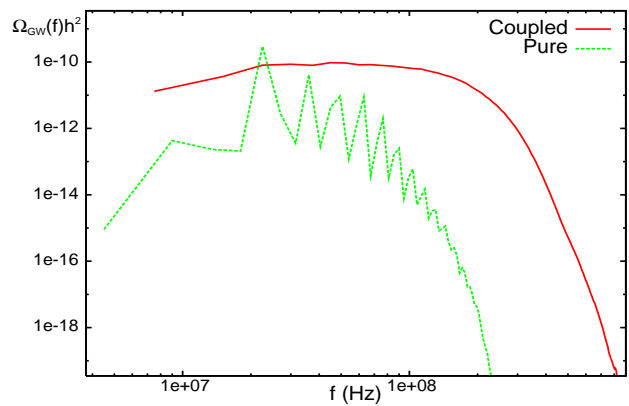}
\end{center}
\vspace*{-5mm}
\caption{Today's ratio of gravitational waves normalized to radiation
energy density, for both the coupled and the pure case. We took 
$g_{*}/g_0 = 100$ to redshift the spectra from the time of the end 
of production till today.}
\label{fig8}
\vspace*{-3mm}
\end{figure}

Of course, in either case, with and without
an extra field $\phi$, in order to predict today's spectral window of
the GW spectrum, we have, first, to normalize their energy density
at the end of GW production to the total
energy density at that moment. Secondly, we have to redshift the GW
spectra from that moment of reheating, taking into account that the
rate of expansion have changed significantly since the end of
inflation, see Eq.(\ref{redshift}). In particular, the shape and 
amplitude of GW spectra for the case with the extra scalar field
coupled to the inflaton with $q = 120$, see Fig.~\ref{fig8}, 
seems to coincide 
with the espectra shown in Ref.~\cite{DufauxGW}. 
On the other hand, we also reproduce in Fig.~\ref{fig8} a similar
spectra to the one shown in~\cite{TkachevGW}, for the case of the pure
quartic model. 
Thanks to the tremendous gain in
computer power, we were able to resolve the 'spiky' pattern
of the spectrum with great resolution. For the first time,
it is clearly observed the exponential tail for large frequencies, see
Figs.~\ref{fig7}, \ref{fig8}, not shown in Ref.~\cite{TkachevGW}.  The most
remarkable fact, is that we also confirm that the peak structure in
the GW power spectrum, see Fig.~\ref{fig7}, remains clearly visible at
times much later than the one at which those peaks have dissapeared in
the scalar fields' power spectrum. So, as pointed out in
Ref.~\cite{TkachevGW}, this characteristic feature distinguish this
particular model from any other.

Let us emphasize that we have run the simulations till times 
much greater than that of the end of the resonance stage, both for 
the pure and the coupled case. The role of the
turbulence period after preheating seems, therefore, not to be very 
important, despite its long duration. Apparently, the \textit{no-go} 
theorem about the suppresion of GW at turbulence, discussed in~\cite{DufauxGW}, 
is fulfilled. In Refs.~\cite{CEWB,Dufaux} it was 
pointed out that gauge couplings or trilinear interactions could be 
responsible for a fast thermalization of the universe after inflation 
(see also Ref.~\cite{FK}), but as long as this takes place after the end 
of the resonace stage, in principle this should not affect the results 
shown above.

\begin{widetext}

\begin{figure}[h]
\begin{center}
\includegraphics[width=10cm,height=16cm,angle=-90]{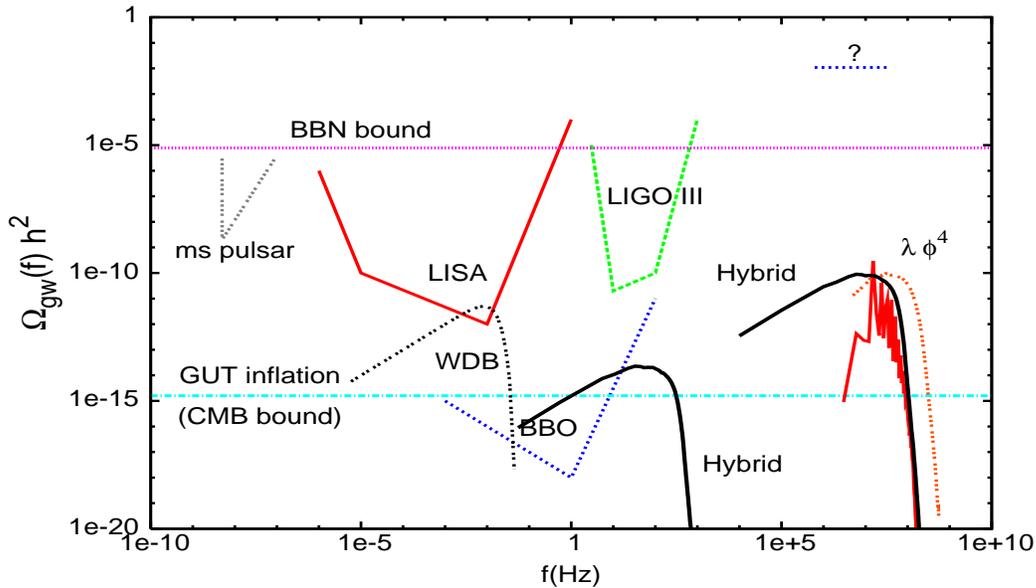}
\end{center}
\vspace*{-5mm}
\caption{The sensitivity of the different gravitational wave
experiments, present and future, compared with the possible stochastic
backgrounds; we include the White Dwarf Binaries (WDB)~\cite{WDB} and
chaotic preheating ($\lambda\phi^4$, coupled and pure) for comparison.
Note the two well differentiated backgrounds from high-scale and
low-scale hybrid inflation. The bound marked (?) is estimated from 
ultra high frequency laser interferometers' 
expectations~\cite{Nishizawa2007}.}
\label{fig5}
\vspace*{-3mm}
\end{figure}

\end{widetext}

\section{Conclusions}

To summarize, we have shown that hybrid models are very efficient
generators of gravitational waves at preheating, in three well defined
stages, first via the tachyonic growth of Higgs modes, whose gradients
act as sources of gravity waves; then via the collisions of highly
relativistic bubble-like structures with large amounts of energy
density, and finally via the turbulent regime (although this effect
does not seem to be very significant in the presence of scalar
sources), which drives the system towards thermalization. These
waves remain decoupled since the moment of their production, and thus
the predicted amplitude and shape of the gravitational wave spectrum
today can be used as a probe of the reheating period in the very early
universe. The characteristic spectrum can be used to distinguish
between this stochastic background and others, like those arising from
NS-NS and BH-BH coalescence, which are decreasing with frequency, or
those arising from inflation, that are flat~\cite{SKC}.
\\

We have plotted in Fig.~\ref{fig5} the sensitivity of planned GW
interferometers like LIGO, LISA and BBO, together with the present
bounds from CMB anisotropies (GUT inflation), from Big Bang
Nucleosynthesis (BBN) and from milisecond pulsars (ms pulsar).  Also
shown are the expected stochastic backgrounds of chaotic inflation
models like $\lambda\phi^4$, both coupled and pure, as well as
the predicted background from two different hybrid inflation models, a
high-scale model, with $v=10^{-3}M_P$ and $\lambda\sim g^2\sim0.1$,
and a low-scale model, with $v=10^{-5}M_P$ and $\lambda\sim g^2\sim
10^{-14}$, corresponding to a rate of expansion $H\sim 100$ GeV. The
high-scale hybrid model produces typically as much gravitational waves
from preheating as the chaotic inflation models. The advantage of
low-scale hybrid models of inflation is that the background produced
is within reach of future GW detectors like BBO~\cite{BBO}. It is
speculated that future high frequency laser interferometers could
be sensitive to a GWB in the MHz region~\cite{Nishizawa2007}, although
they are still far from the bound marked with an interrogation sign.

For a high-scale model of inflation, we may never see the predicted GW
background coming from preheating, in spite of its large amplitude,
because it appears at very high frequencies, where no detector has yet
shown to be sufficiently sensitive. On the other hand, if inflation
occured at low scales, even though we will never have a chance to
detect the GW produced during inflation in the polarization
anisotropies of the CMB, we do expect gravitational waves from
preheating to contribute with an important background in sensitive
detectors like BBO. The detection and characterization of such a GW
background, coming from the complicated and mostly unknown epoch of
rehating of the universe, may open a new window into the very early
universe, while providing a new test on inflationary cosmology.
\\

\section*{Acknowledgments} 
We wish to thank Andr\'es D\'iaz-Gil, Jean-Francois Dufaux, Richard
Easther, Gary Felder, Margarita Garc\'\i a P\'erez, John T. Giblin
Jr., Seiji Kawamura, Lev Kofman, Andrei Linde, Eugene A. Lim, Luis
Fernando Mu\~noz-Mej\'{\i}as and Mischa Sall\'e for very useful
comments, suggestions and constructive criticism. This work is
supported in part by CICYT projects FPA2003-03801 and FPA2006-05807,
by EU network "UniverseNet" MRTN-CT-2006-035863 and by CAM project
HEPHACOS S-0505/ESP-0346. D.G.F. and A.S. acknowledges support from a
FPU-Fellowship from the Spanish M.E.C.  We also acknowledge use of the
MareNostrum Supercomputer under project AECT-2007-1-0005.

\end{document}